# A Reference Model for Information Quality in an IT Governance Context

Dirk Steuperaert[a,b]*, Geert Poels[a,c] and Jan Devos[a]

[a]*Business Informatics and Operations Management Department, Faculty of Economics and Business Administration, Ghent University, Belgium*

[b]*Management Information Systems Department, Department MIS Antwerp University, Belgium;*

[b]*CVAMO core lab Flanders Make @Ghent University;*

Dirk Steuperaert, dirk.steuperaert@ugent.be, https://orcid.org/0000-0002-7009-4183

Dirk Steuperaert is currently a PhD candidate at Ghent University and Antwerp University. His research focuses on IT Governance Systems and COBIT, and he is also lecturing in this field. Dirk has over 35 years of experience in IT, IT Audit, and IT Governance in the financial sector and in consulting. Dirk is a major contributor to the development of various ISACA (Information Systems Audit and Control Association), including COBIT.

Geert Poels, geert.poels@ugent.be, https://orcid.org/0000-0001-9247-6150

Geert Poels is a professor of Business Informatics at Ghent University. He teaches intermediate and advanced courses on Information Systems, Enterprise Architecture, and Information Technology Management. His research covers a broad area, including Conceptual Modelling, Ontology, Requirements Engineering, Privacy and Personal Data Protection, and IT Governance. He was a co-developer of the COBIT 2019 Framework.

Jan Devos, jang.devos@ugent.be, https://orcid.org/0000-0001-7408-6853

Jan Devos is a professor of Information Technology and Information Systems at Ghent University. His current research interests are IT governance in SME's, design science, IS failures, and IT Security. Prof. Devos has published several articles on IT and SMEs and was a speaker at many international academic and business conferences.

# A Reference Model for Information Quality in an IT Governance Context

IT Governance systems are increasingly required to keep today's organizations functioning. IT Governance requires a holistic system of interacting components, including processes, organizational structures, information, and others. Performance management of IT Governance systems is of utmost importance to maintain their effectiveness. Capability models are used to assess and manage IT Governance process performance, whereas similar mechanisms are lacking for other types of IT Governance system components, e.g. information. In this paper, we focus on how to define the quality of IT Governance information, as a proxy for the performance of the information component of the IT Governance system. Using a Design Science approach, we iteratively develop, based on theory, and empirically evaluate, based on expert validation, a reference model for IT Governance information quality, i.e., the Information Quality Reference Model (IQRM) that can be used for assessing the quality of IT Governance information items. The model is comprehensive yet manageable and provides a basis for building a capability model for IT Governance information.

**Introduction**

Today's world has become heavily reliant on information and information technology (IT) in all aspects of life, including personal, government, and business. At the same time, there is a constant stream of reported IT incidents, such as cyber-attacks, system failures, data breaches, and failed projects, and there is likely an even greater number of lower-profile incidents that go unreported or don't make headlines. All these incidents negatively impact the overall performance of IT, and therefore the achievement of business objectives using IT for the affected organizations. To address the performance and contribution of IT, many organizations have put in place an IT Governance system, which is a collection of processes, organizational structures, policies, information flows and other mechanisms to direct and monitor how IT is providing value to the organization (Peterson, 2004). That governance system itself also needs to perform well to realize the benefits of IT Governance. The question on performance management of a governance system is relevant because it has been demonstrated that good performance of the IT Governance system is strongly correlated to IT performance (De Haes &Van Grembergen, 2009, 2020, Weill & Ross, 2004).

IT Governance scholars (Dehaes & Van Grembergen, 2009) have identified the interrelated components of an IT Governance system as interacting processes, organizational structures and relational mechanisms. In more recent research and trade publications this three-layer structure of an IT Governance system is extended with more components. The prevailing practitioner IT Governance Framework COBIT (ISACA, 2018) argues that an IT Governance system for an organization should be built on seven components, i.e. processes, organizational structures, culture and behavior, information items or flows, people and skills, infrastructure and principles, and policies and procedures. Hence, performance assessment and management of an IT Governance system should consider the performance of all its interrelated components.

Process capability models are a well-established technique to assess the performance of processes (Van Looy et al., 2013) and can thus be used to manage the performance of the process component of an IT Governance system. However, there is limited research available on the measurement and management of the performance of the other components of an IT Governance system. The viability and potential utility of a capability model for organizational structures has already been demonstrated (Steuperaert et al., 2021). Still, there remain several other components of an IT Governance system for which there is currently no performance assessment model.

In this paper we focus on one such component for which there is no performance assessment model yet: the information items that are required to make an IT Governance system perform. So, we investigate in this paper how we can facilitate the assessment of the performance of the information items that are required for effective IT Governance. The performance of information is not a known concept or research abstraction, but by logical deduction, it follows that the performance of information in an IT Governance context means that the information contributes to the objectives of the IT Governance system. In other words, information is performing well when the information is fit for the purpose of contributing to the effectiveness of the IT Governance system.

The ISO 9001:2015 standard defines fitness for purpose as a quality dimension (ISO, 2015). Information quality is a concept that has been researched before though not in the context of IT Governance. While our goal is to develop a capability model that can be used to assess and manage the quality of the information items required for an effective IT Governance system, the first research objective is to develop a *reference model for IT Governance information quality*. As a building block of the envisioned capability model, this reference model breaks down the quality of IT Governance information into dimensions and sub-dimensions and provides precise definitions of the information quality attributes situated at these (sub-)dimensions. It is furthermore argued that understanding attributes of information contributes to the effective use of information for its goals (Boell & Cecez, 2010), which is another argument for decomposing the concept 'information quality' into a set of attributes or dimensions.

Although comprehensive quality models for information can be found – mainly in trade journals or in professional associations' frameworks – they are not specific to IT Governance information. For instance, the information quality model proposed by

DAMA (2017) is geared towards data (structured, to be used in databases) rather than towards information. Furthermore, these models are not validated in an IT Governance context. This constitutes the research gap that we address in this paper. Commercial (consulting) organizations have also developed models on data quality (Gartner, 2007)

For the reasons explained above, our research objective is to develop a reference model for IT Governance information quality, which we will call the *Information Quality Reference Model* (IQRM). It would constitute a reference that can be used to express and assess the quality of IT Governance information items and as such can form the basis for a capability model for IT Governance information.

The context and problem statement above lead us to formulate our main research question (RQ1).

**RQ1: Which attributes of information constitute a practical and comprehensive reference model for information quality in the context of IT Governance?**

The answer to RQ1 will deliver the IQRM. As the IQRM will be used in future research as the basis for developing a capability model for IT Governance information, we also formulate a second research question (RQ2).

**RQ2: To what extent is the importance of these information quality attributes dependent on the purpose the information serves in an IT Governance system?**

Research question RQ2 aims to investigate whether there are contingency factors in an IT Governance system that would require differentiating between IT Governance information items when it comes to assessing their quality based on the IQRM. The answer to this question is an important step towards the development of a capability model for IT Governance information.

The remainder of this paper is structured as follows: In section 2 we review related work. In section 3 we discuss the overall methodology and approach. In section 4 we develop and validate the IQRM. In section 5 we conclude and discuss the limitations of our research, as well as potential future research.

**Related Work**

The academic literature does not offer a comprehensive or generally accepted framework for information quality in the context of IT Governance. Further, papers on information governance do not discuss information quality. However, we found papers that present conceptual models of data or information quality independent of the context of the use of the data or information. These models provide a starting point for designing the IQRM, hence the review of these models is presented in section 4 where we describe the development of the IQRM.

The grey literature provides standards and frameworks for data or information management, some of which deal with quality. The main practitioner organization on data management – DAMA International – has developed a body of knowledge on data management called DMBOK, which in its original version included a section that discussed data quality and presented the following set of six core data quality dimensions (Table 1).

| Data Quality dimensions according to DAMA UK | |
|---|---|
| **Accuracy** | Accuracy measures how well the available data corresponds with experiences in the real world. |
| **Completeness** | Completeness covers the extent that data and its metadata are present |
| **Validity** | Validity confirms that data behaves according to business expectations |
| **Consistency** | Consistency describes how similar the original data and that delivered to another system, storage, interface, or through a pipeline match |
| **Integrity** | Integrity measures how well any data set maintains its structure and relationships after data processes execute |
| **Uniqueness/Deduplication** | This dimension uncovers one or more versions of an entity described by the data |

Table 1 - Data Quality Dimensions as per DMBOK (DAMA, 2017)

Additional characteristics of data, listed by DMBOK, include the following (DAMA, 2017, pp. 457-458):

- "Usability: Is the data understandable, simple, relevant, accessible, maintainable and at the right level of precision?
- Timing issues (beyond timeliness itself): Is it stable yet responsive to legitimate change requests?
- Flexibility: Is the data comparable and compatible with other data? Does it have useful groupings and classifications? Can it be repurposed? Is it easy to manipulate?
- Confidence: Are Data Governance, Data Protection, and Data Security processes in place? What is the reputation of the data, and is it verified or verifiable?
- Value: Is there a good cost / benefit case for the data? Is it being optimally used? Does it endanger people's safety or privacy, or the legal responsibilities of the enterprise? Does it support or contradict the corporate image or the corporate message?"

Both lists above were then synthesized in the 2nd edition of the body of knowledge (i.e., DMBOK2) into a set of eight 'common dimensions' of data quality (Table 2).

| Dimension of Quality | Description |
| --- | --- |
| Accuracy | Accuracy refers to the degree that data correctly represents 'real-life' entities. |
| Completeness | Completeness refers to whether all required data is present |
| Consistency | Consistency can refer to ensuring that data values are consistently represented within a data set and between data sets, and consistently associated across data sets. It can also refer to the size and composition of data sets between systems or across time. |
| Integrity | Data Integrity (or Coherence) includes ideas associated with completeness, accuracy, and consistency. |
| Reasonability | Reasonability asks whether a data pattern meets expectations |
| Timeliness | The concept of data Timeliness refers to several characteristics of data. Measures of timeliness need to be understood in terms of expected volatility – how frequently data is likely to change and for what reasons. |
| Uniqueness/Deduplication | Uniqueness states that no entity exists more than once within the data set |
| Validity | Validity refers to whether data values are consistent with a defined domain of values. A domain of values may be a defined set of valid values (such as in a reference table), a range of values, or value that can be determined via rules. |

Table 2 - Data Quality Dimension as per DMBOK2 (DAMA, 2017)

One notices that the definition of Integrity refers to three other dimensions, indicating potential overlap or ambiguity. In addition, the detailed definitions in DMBOK2 for the dimensions of Validity and Reasonability seem to be very similar.

Whereas DAMA has published its set of common data quality dimensions as part of DMBOK2, referring to multiple academic publications that support their model, different and additional quality dimensions can be found in those references. We interpret this as an indication that the model proposed by DAMA has not reached a sufficient level of completeness yet.

For instance, we found that the 'data management wiki' from the DAMA NL chapter (Black & Van Nederpelt, 2020) defines 44 data quality dimensions. This set of quality dimensions is more complete than the one described above, but there might be concern about potential overlap or duplication. Another concern would probably be the practicality of having to deal with such many dimensions. We believe that a workable model should therefore contain fewer dimensions than the ones proposed in this datawiki.

Redman (1996) identified also multiple data quality dimensions. His model described the dimensions of the data model, data values and data representation. The latter two seem to be most relevant to our research and include Accuracy, Completeness, Currency and Consistency for data values, and Appropriateness, Interpretability, Portability, Format precision, Format flexibility, Ability to represent null values, Efficient use of storage, and Physical instances of data being in accord with their formats for data representation.

Finally, Myers (Myers, 2015) has, based on the work of scholars, synthesized the quest for generally agreed data quality dimensions, and has proposed a model (the Conformed Dimensions of Data Quality) with 11 data quality dimensions and 47 sub-dimensions. Since timewise, this research ran in parallel with our study, we have included a comparison between his results, DMBOK2 and our results in Section 5.

Without claiming exhaustiveness, we have found that many articles in the grey literature are not specific on what exactly data quality means, but rather elaborate on good data management practices (Montecarlodata, 2023; Performance Improvement Council, 2016). Whereas management practices for IT Governance information is a relevant topic for the development of a capability model, such practices fall outside the scope of the IQRM.

**Methodology**

The research paradigm guiding our investigation is Design Science (Hevner, 2004, Peffers et al., 2007). We start from a problem that is relevant to practice and for which there is no solution (i.e., lack of a reference model for IT Governance information quality). We build the solution in the form of an artifact (i.e., the IQRM) using a rigorous approach that builds upon knowledge available in the literature that we adapt to our problem context (i.e., performance management of IT Governance systems). We evaluate the solution by validating the artifact's design based on the domain knowledge of experts. In practice, we have performed the following four research activities (Figure 1):

- Initial Artefact Design: The initial design of the artefact was based on a literature study on information governance and information quality that was performed as part of a project for the ISACA organization. Comparing and synthesizing existing information quality models led to the proposal of an information quality model to be used in the context of IT Governance. This model was not empirically validated but consensually agreed on and further refined by a small group of IT Governance experts associated with ISACA. It was then used as the Information Reference Model (IRM) in the COBIT 5 Framework for IT Governance (ISACA, 2012).

- Second Iteration of the Artefact Design: Before validating the artefact design, we updated the literature study that was the basis for the initial design. This was needed given that many years had passed since the publication of the IRM as part of the COBIT 5 Framework. Based on a peer review of the results of the additional literature study we updated the initial design to obtain a second iteration design of the artefact.

- Validation of the Artefact Design: We composed a large and diverse expert panel (n = 35) to validate the second iteration design of the artefact. The domain knowledge of experts was used to evaluate the structure (i.e., dimensions and sub-dimensions) and content (i.e., definitions of quality attributes situated at these (sub)-dimensions) of the proposed reference model. They were also asked about contingency factors in IT Governance systems that would impact the importance of the quality attributes.

- Final Artefact Design: based on an analysis of the feedback from the expert panel, we proposed a final design which is what we present as the IQRM.

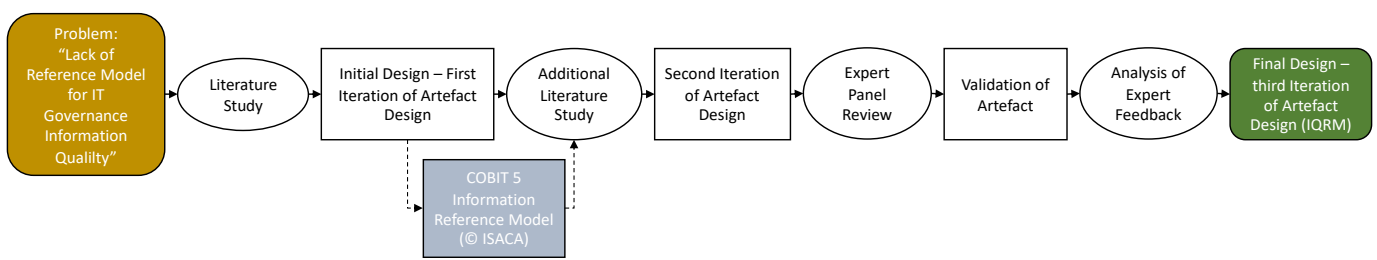

Figure 1 - Methodology Overview

*Initial Artefact Design*

An exploratory review of the academic literature that we conducted before the start of this study indicated an absence of studies that investigated desirable attributes of information required for IT Governance. We found no model of information quality specific to this context. Also, literature on the governance of information, as different from the governance of information systems or IT, was not abundant. The most elaborate governance model we found was that of Khatri and Brown (Khatri & Brown, 2010). This model merely emphasizes quality as a governance domain and provides just a few examples of data quality dimensions: Accuracy, Timeliness, Completeness and Credibility.

We therefore decided to conduct a Focused Literature Review (Webster & Watson, 2002) on models of information quality in general. The Focused Literature Review method differs from the Systematic Literature Review method (Kitchenham et al., 2009). Whereas a systematic literature review strives for exhaustiveness in its review of the state-of-the-art on a particular research topic, a focused literature review highlights a disciplinary perspective on a research topic, which translates into a more selective choice of sources. It also has a more quality-oriented focus than a systematic literature review.

The disciplinary perspective we chose was that of Management Information Systems, so excluding, for instance, studies on the quality of data in other contexts like signal processing, computer networks or machine learning. The quality-oriented focus translated into a literature search that started from the most authoritative sources on information quality, which were in the period that we performed this literature review (2010-2011), professors Stuart Madnick and Richard Wang of MIT. So, we looked up the work of these authors and identified other papers using forward and backward snowballing.

The initial design of the reference model for IT Governance information quality was then based on a synthesis and comparison of the proposed models of information quality we found in the selected papers. Since the model of Wang and Strong (1996) was published in what seemed to be the seminal paper on information quality, we adopted this model as our main theoretical foundation but adapted it where relevant based on more recent models that we found. Most of these models were derived from the model of Wang and Strong, with some exceptions like the model of Nelson, Todd and Wixom (2005). Through this process, we used relevance to the IT Governance context and a focus on information rather than data as guidelines.

*Second Iteration of the Artefact Design*

The initial artefact design was the basis for the COBIT 5 IRM as published by ISACA (ISACA, 2013). A Google Scholar search for "Information Reference Model" AND "COBIT" did not identify any studies that researched or reviewed the IRM since its publication in 2013. Therefore, we performed an additional literature study to align the initial design of our artefact to the latest insights and research results. For that reason, additional literature reviews were performed.

A new literature study, reviewing a total of 71 papers, was performed by Desplenter (2019) as part of another research project. Several enhancements to the COBIT 5 IRM were suggested, including new information quality criteria, existing quality criteria that can be merged, and the definition of information quality sub-criteria. We used the results of this literature study for the second iteration design of the artefact.

This was followed by another round of literature review searching for papers on "Information Quality Management". The review period was 2012 – 2022. Most of the papers found heavily focus on data management or data governance practices, but seldom or not elaborate on the definition of information quality. Furthermore, none of the papers discussed information quality in the context of IT Governance systems. Since we have not identified any more information quality attributes based on this part of the literature research, the results are not further discussed in the paper.

*Validation of the Artefact Design*

*Expert Panel Composition*

The second iteration design of the envisioned reference model was submitted for validation to an expert panel. To ensure the success of the expert validation, it was important to assemble a qualified expert panel and to design a well-crafted questionnaire. The criteria used to select the expert panel, where each individual panel member should satisfy the first two selection criteria, included the following:

- Significant relevant professional experience (with a preference for highly experienced individuals and a limited number of junior professionals), because we want the views of people with adequate experience levels.
- Relevant expertise in either information governance, data management, academic research in information management, or IT Governance information usage. Since our research is situated in the IT Governance domain and is related to managing and using the information flows of IT Governance systems, we looked for this expertise in the academic community, in the community of IT Governance professionals, and in the consulting world. To compose a diverse panel of experts, we were interested in the views of people with different professional backgrounds.
- International representation - we sought the input of international experts to reduce any local biases.

We invited 63 experts to participate in the validation and received 35 positive answers.

The panel was composed of four categories of experts as shown in Table 3, where we also include the selection methods for each category, as well as the number and percentage of respondents for each category.

| Category | Selection Method | # respondents |
|---|---|---|
| **Academia** | We invited members of local and associated research groups on IS systems and IT Governance | 4 (11%) |
| **Experienced Governance Professionals** | We invited alumni and current students from executive master programs in Information Systems Governance at the Antwerp Management School. | 17 (49%) |
| **Experienced Consultants** | We invited experienced consultants with an IT Governance, IT Security, Data Management or Data Governance background. | 10 (29%) |
| **Information Users** | Experienced IT Governance information users from various non-IS backgrounds (Law, Education, Manufacturing Industry) | 4 (11%) |

Table 3 - Expert Panel for Information Quality Criteria Validation

The composition and specific relevant demographic and professional characteristics of the panel members are included in Appendix 1.

*Questionnaire submitted to the panel*

The questionnaire aimed to obtain the expert panel's opinions on information quality, and to verify the consistency, completeness, and validity of the proposed reference model. For that purpose, our questionnaire contained five questions, as shown in Appendix 2. None of these questions referred specifically to the use of information as part of IT Governance systems as we assumed that the experts would consider this context given their professional background and experiences.

The first question was designed to solicit the unbiased opinion of the experts on what constitutes information quality for them, in their own words. The question was open, and we did not mention the existence or any specifics of the reference model yet, so we would receive unfiltered answers. The benefit of this question was twofold, i.e. (a) it would allow us to identify any potentially missing information quality attributes of the model, and (b) it would support the validation of the proposed model in the most unbiased way possible.

*Q1. When you consider 'information quality', what are the criteria you would use to decide whether the information has quality or not? (Information can be any type of information or data)*

Every participant responded to this question, and a comprehensive compilation of their responses is available in Appendix 3. We analyzed their answers as explained in Table 4.

| Step 1 | The respondents' answers are matched literally against the information quality criteria in the second iteration design of the reference model, and a frequency table of matched answers is compiled. (coding system applied: literal match with any of the quality attributes in the reference model) |
|---|---|
| Step 2 | The unmatched respondents' answers were analyzed separately by each author of the paper and when found to be a synonym of, or an equivalent to, one of the criteria in the reference model, matched to these criteria by updating the frequency table. (coding system applied: synonym of any of the quality attributes in the reference model, peer reviewed) (See Appendix 4 for details) |
| Step 3 | The remaining respondents' answers are further analyzed and identified as potential new criteria for inclusion in the reference model or are classified as 'not a relevant information quality criterion', with ample, co-author independently reviewed justification. (See Appendix 5 for details) |

Table 4 - Analysis Steps for Open Answers to Q1

The second question presented the panel with the 18 information quality criteria of the second iteration design of the reference model and probed for the relevance of each of the presented information quality criteria. We asked the panelists to rate each information quality criterion on a three-value Likert scale (Relevant – Somewhat Relevant – Not Relevant) and to provide us with any comments they would have.

*Q2. To what extent do you think each of the following 18 criteria would be relevant when assessing the quality of information, regardless of the purpose of the information?*

In addition to creating frequency tables of the answers to this question, we also assessed the consistency of the answers between questions 1 and 2 by comparing the frequency table of the criteria in Q1 (after analysis step 2) with the frequency table of all 'Relevant' answers for each of the information quality criteria in Q2. We would expect to find a positive correlation between these two tables. Indeed, if certain criteria are mentioned frequently in response to the open question (i.e., Q1), it indicates that respondents spontaneously recognize their importance. Consequently, these criteria are expected to receive higher relevance scores in response to the closed question (i.e., Q2).

We then asked whether any contingencies for the importance of information quality criteria could be identified in the third question. The main reason is to establish whether our assumption that there are potential contingency factors in IT Governance systems for the relevance of information quality criteria to certain information items, is justified and hence would require further investigation. This information was important to us in the context of our goal to develop, based on the envisioned IQRM, a capability model for IT Governance information.

*Q3. Can you think of any examples where the importance of quality criteria depends on the purpose the information is serving? <u>Example</u>: timeliness will be much more important for a financial transaction compared to a yearly management plan.*

The analysis of the answers to this question will mainly consist of observing the number of potential contingencies reported. In the second order, a qualitative comparison will be made with the answers to Q2 and the frequency tables obtained.

Finally, we asked the participants – after they had been exposed to the full list of criteria – whether they believed any additional criteria could be relevant for assessing information quality (Question 4), and if so, which ones they would identify (Question 5)

*Q4. Would you see any additional criteria that you think are relevant for defining and assessing information quality?*

*Q5. Which additional criteria according to you are relevant for information quality?*

The analysis of the answers to these questions would consist of compiling a frequency table of how many additional criteria could be identified, and a critical analysis of the suggested additional criteria with a decision to add or not add them to the reference model.

*Final Artefact Design*

Based on an analysis of the feedback from the expert panel, adjustments to the second iteration design of the artefact were made. Adjustments included removing quality criteria, adding criteria or modifying the definition of criteria. Adjustments were peer-reviewed by the research team.

**Results**

*First Iteration or Initial Artefact Design*

The guidance provided by information governance models (e.g., Khatri & Brown, 2010) is very much focused on accountabilities and responsible roles, and on the performance of specific activities or taking of decisions, all of which can be integrated into an IT Governance process reference model with specific objectives and processes for data governance and data management. The scarcity of research on information governance and the high-level nature of the information governance models found, lead to the conclusion that the theoretical foundation of the design of a reference model for IT Governance information quality should be provided by information quality models. The information quality models selected from the literature (i.e., Wang & Strong, 1996; Strong et al., 1997; Kahn et al., 2002; Lee et al., 2002; Nelson et al., 2005) were more concrete regarding the nature and desirable properties of information itself, and thus offer a more suitable theoretical foundation for the initial artefact design.

What both literature streams (i.e., information quality and information governance) had in common was the lack of differentiation of the concepts of data and information. While theoretically data and information are different, most studies used these terms interchangeably, except for a certain tendency to refer to information as 'data in use' and to data proper as 'data residing in systems' or 'data being processed' *(*Madnick et al., 2009). The subtle difference found in some publications did not justify making a hard distinction, hence we decided that the envisioned IQRM should refer to information as including data.

A comparative analysis of the selected information quality models converged to a relatively stable set of about 15 quality attributes in four distinct dimensions. Although the exact selection and definition of attributes differed amongst studies as did the naming of the four dimensions, there seemed to be consensus on the different perspectives on information quality. The four information quality dimensions found in the literature are:

- *Intrinsic quality* is a property of information that refers to its informational content, independent of its purpose or use or for whom the information is intended. A representative quality attribute for this dimension is Accuracy, which refers to how well the information as an observation or measurement matches the actual phenomenon or object being observed or measured. Other attributes in this dimension can be characterized as related to the reliability of the information (e.g., Believability, Objectivity, Reputation).
- *Contextual quality* adds a dimension of information use to information quality. As opposed to intrinsic quality, this dimension considers information quality relative to the (intended) use that is made of the information. This dimension can also be characterized as the information's fitness for purpose which needs to be evaluated in the context of this purpose. Often mentioned attributes of contextual quality are Usefulness, Relevance, and Timeliness. While intrinsic quality impacts contextual quality, it is itself independent of contextual quality. For instance, inaccurate information will negatively impact perceived usefulness but information with no identified use can still be accurate or precise.

- *Representational quality* is a dimension that also impacts contextual quality or could be considered a sub-dimension of contextual quality. This dimension refers to the format of the information and not to its content as with intrinsic quality. So, this dimension captures how the way information is represented impacts (perceptions of) contextual quality, with attributes like Understandability, Conciseness and Consistency of representation, and Ease of manipulation.

- Finally, *accessibility* is distinguished as an orthogonal dimension of information quality that is related to who is (allowed) accessing the information. This dimension is thus independent of the other three dimensions and allows conceptualizing information security as a quality property. Attributes can be formulated positively (e.g., Availability) or negatively (e.g., Restricted access).

Based on the comparative analysis of the selected information quality models, the initial design of the reference model was constructed and deliberated by IT Governance experts from ISACA who then included it as the IRM in the COBIT 5 Framework (ISACA, 2013). The same model is referenced, but not further explained, in the later COBIT 2019 Framework (ISACA, 2018).

The IRM, our first design iteration of the IQRM, is shown in Table 5. The model distinguishes three information quality dimensions with in total 15 information quality criteria.

| Information Quality Reference Model – First Design Iteration | | |
|---|---|---|
| Quality Dimension | Quality Attributes | |
| <u>Intrinsic</u><br>The extent to which data values are in conformance with the actual or true values | Accuracy | The extent to which information is correct and reliable |
| | Objectivity | The extent to which information is unbiased, unprejudiced and impartial |
| | Reputation | The extent to which information is highly regarded in terms of its source or content |
| | Believability | The extent to which information is regarded as true and credible |
| <u>Contextual</u><br>The extent to which information is applicable to the task of the information user and is presented in an intelligible and clear manner, recognizing that information quality depends on the context of use | Relevancy | The extent to which information is applicable and helpful for the task at hand. |
| | Completeness | The extent to which information is not missing and is of sufficient depth and breadth for the task at hand |
| | Currency | The extent to which information is sufficiently up to date for the task at hand |
| | Appropriate Amount | The extent to which the volume of information is appropriate for the task at hand |
| | Concise Representation | The extent to which information is compactly represented |
| | Consistent Representation | The extent to which information is presented in the same format |
| | Interpretability | The extent to which information is in appropriate languages, symbols and units, and the definitions are clear |
| | Understandability | The extent to which information is easily comprehended |
| | Ease of Use | The extent to which information is easy to manipulate and apply to different tasks |
| <u>Security</u><br>The extent to which information is available or obtainable | Availability | The extent to which information is available when required, or easily and quickly retrievable |
| | Restricted Access | The extent to which access to information is restricted appropriately to authorized parties |

Table 5 – Information Quality Reference Model – First Design Iteration (ISACA, 2013; 2018a)

*Second Iteration*

The update of the literature review for the period 2012-2019 delivered a few useful results, based on which we made changes to the initial design. The design decisions were based on the following workflow. First, an extensive literature review was held which identified 71 relevant publications on data and information quality (Desplenter, 2019). From these publications, a list of 90 data quality dimensions was identified. In the next phase, a frequency table was constructed and only those quality dimensions that were mentioned at least five times across all publications were retained, resulting in

36 dimensions remaining. In the final phase, a qualitative review for dimensions that were too hard to measure, incorporating other dimensions or being synonymous to other dimensions was held to further reduce the number of quality dimensions to 25. Because of this still relatively large number, intermediate levels of grouping were also introduced.

This activity was followed by a subsequent critical review by the research team. The team concluded on the one hand that the resulting number of information quality criteria is (too) high to work with in practice, and on the other hand that the remaining 25 quality dimensions contain inconsistencies that can be traced back to interpretations of the data collected in the literature study. However, the literature study also provided useful suggestions to incorporate in the design of the artefact. The outcome of the review of the literature study's results constitutes the second iteration of the design of the information quality reference model, as shown in Table 6.

| Information Quality Reference Model – Second Design Iteration ||||
|---|---|---|---|
| **Quality Dimension** | **Quality Sub-Dimension** | **Quality Attributes and Sub-Attributes** ||
| <u>Intrinsic</u><br>The extent to which data values are in conformance with the actual or true values | <u>Correctness</u><br>The extent to which the data is accurate and precise | Accuracy | The extent to which information is semantically correct |
| | | Precision | The extent to which information is syntactically correct |
| | <u>Reliability</u><br>The extent to which information is believable, reputed highly and objective | Reputation | The extent to which information is highly regarded in terms of its source or content |
| | | Objectivity | The extent to which information is unbiased, unprejudiced and impartial |
| | | Traceability | The extent to which information's source can be found |
| <u>Contextual</u><br>The extent to which information is applicable to the task of the information user | <u>Adequacy</u><br>The extent to which information is sufficient for the task at hand | Appropriate Amount | The extent to which the volume of information is appropriate for the task at hand |
| | | <u>Usability</u><br>The extent to which information is easily manipulated and efficiently used | Ease of Use: The extent to which information is easy to manipulate and apply to different tasks |
| | | | Efficiency: The extent to which information is able to quickly meet the information needs for the task at hand |

| Information Quality Reference Model – Second Design Iteration ||||
|---|---|---|---|
| **Quality Dimension** | **Quality Sub-Dimension** | **Quality Attributes and Sub-Attributes** ||
| | | Completeness | The extent to which information is not missing and is of sufficient depth and breadth for the task at hand |
| | | Relevancy | The extent to which information is applicable and helpful for the task at hand. |
| | | <u>Time</u><br>The extent to which information is timely and current | Currency — The extent to which information is sufficiently up to date for the task at hand |
| | | | Timeliness — The extent to which the age of the data is appropriated for the task at hand |
| | <u>Representation</u><br>The extent to which information is presented in an intelligible and clear manner | Interpretability | The extent to which information is in appropriate languages, symbols and units, and the definitions are clear |
| | | <u>Clarity</u><br>The extent to which information is consistently and concisely presented | <u>Consistency</u> — The extent to which information is presented in the same format |
| | | | <u>Conciseness</u> — The extent to which information is compactly represented |
| | | Understandability | The extent to which information is easily comprehended |
| <u>Security</u><br>The extent to which information is available and accessible | Accessibility | | The extent to which access to information is restricted appropriately to authorized parties |
| | Availability | | The extent to which information is available when required, or easily and quickly retrievable |

Table 6 – Information Quality Reference Model – Second Design Iteration

In this second iteration, we added additional grouping levels to improve the hierarchical structure of the model. The 'intrinsic' dimension was split into 'correctness' and 'reliability' sub-dimensions, each containing a set of attributes. Also, the 'contextual' dimension was split into an 'adequacy' and 'representation' sub-dimension, each containing different attributes which in some cases were now positioned as sub-attributes of other quality attributes. Ease of Use was now defined as a sub-attribute of a new Usability attribute. Likewise, Currency was defined as a sub-attribute of a new Time attribute. Finally, Consistency and Conciseness were defined as sub-attributes of a new clarity attribute. Attributes that have sub-attributes were defined such that when an information item exhibits the quality defined by all sub-attributes, it exhibits the quality defined by the attribute.

We further added new quality attributes or sub-attributes of quality attributes already present in the initial design. Precision was added as a quality attribute next to Accuracy in the 'correctness' sub-dimension of the intrinsic quality dimension. Traceability was a new quality attribute in the 'reliability' sub-dimension of the intrinsic quality dimension, where it replaced Believability. Efficiency was added next to Ease of Use as a sub-attribute of Usability. Likewise, Timeliness was added next to Currency as a sub-attribute of Time.

Minor changes to the model included renaming Restricted Access as Accessibility and modifying the definition of Accuracy such that it more clearly belongs to the 'correctness' sub-dimension of intrinsic quality as in the initial design Accuracy also referred to 'reliability' which is another sub-dimension of intrinsic quality.

*Artefact validation*

The survey with the expert panel ran between April 27$^{th}$ and May 15$^{th}$, 2022. Thirty-five responses were received with complete answers for the first two questions.

*Q1 – When you consider 'information quality', what are the criteria you would use to decide whether the information has quality or not? (N=35)*

Table 7 contains a frequency table of how many experts mentioned each of the 18 lowest-level quality attributes of the second iteration design of the reference model (i.e., the sub-attributes or attributes without sub-attributes of Table 6) in response to the open question Q1. The frequencies were determined according to the analysis steps 1 and 2 described in Table 4.

| Information Quality Criteria | #Mentions after analysis step 1 | #Mentions after analysis step 2 |
|---|---|---|
| Accuracy (The extent to which information corresponds to the true value) | 14 | 24 |
| Precision (The extent to which different measurements of the information are close to each other) | 1 | 6 |
| Reputation (The extent to which information is highly regarded in terms of its source or content) | 0 | 17 |
| Objectivity (The extent to which information is unbiased, unprejudiced and impartial) | 3 | 10 |
| Traceability (The extent to which information's source can be found) | 2 | 12 |
| Appropriate Amount (The extent to which the volume of information is appropriate for the task at hand) | 0 | 2 |
| Ease of Use (The extent to which information is easy to manipulate and apply to different tasks) | 1 | 4 |
| Effectiveness (The extent to which information is able to quickly meet the information needs for the task at hand) | 1 | 4 |
| Completeness (The extent to which information is not missing and is of sufficient depth and breadth for the task at hand) | 13 | 13 |
| Relevancy (The extent to which information is applicable and helpful for the task at hand) | 10 | 10 |
| Currency (The extent to which information is sufficiently up to date for the task at hand) | 8 | 9 |
| Timeliness (The extent to which the age of the data is appropriated for the task at hand) | 3 | 4 |
| Interpretability (The extent to which information is in appropriate languages, symbols and units, and the definitions are clear) | 2 | 7 |
| Consistency (The extent to which information is presented in the same format) | 3 | 7 |
| Conciseness (The extent to which information is compactly represented) | 1 | 2 |
| Understandability (The extent to which information is easily comprehended) | 4 | 8 |
| Accessibility (The extent to which access to information is restricted appropriately to authorized parties) | 3 | 3 |
| Availability (The extent to which information is available when required, or easily and quickly retrievable) | 6 | 6 |

Table 7 – Analysis of information quality criteria based on the answers to open question Q1

We observe that every information quality criterion was mentioned by at least two experts, although frequencies differed greatly, varying from 2 mentions (Appropriate Amount, Conciseness) to a maximum of 24 mentions (Accuracy). A few criteria, like Reputation (0 mentions in step 1, 17 after step 2) and Traceability (2 mentions after step 1, 12 after step 2), were not or hardly mentioned literally (i.e., analysis step 1) but had a high number of mappings considering synonyms or equivalent meanings (i.e., step 2 – see Appendix 4). This could indicate that the labelling and short explanation given in the definitions of the information quality criteria were not adequate or not in line with how the experts would call and define them. Referring to the two examples given above, especially the term 'source' was mentioned very frequently in the expert's comments whereas it was absent in our original labels and descriptions of the criteria. These labels and definitions will be subject to possible revision for the final design of the IQRM.

In step 3 of the analysis, no further additions or deletions were needed to the model based on this stage of the validation. The details of this analysis can be found in Appendix 5, where all decisions are explained and justified.

*Question 2–- To what extent do you think each of the following 18 criteria would be relevant when assessing the quality of information, regardless of the purpose of the information? (N=35)*

The results of this question are shown in Table 8. Each criterion could be given one of three ratings on a Likert-scale, i.e. 'Relevant', 'Somewhat Relevant' and 'Not Relevant'. We consider this scale to be equidistant, so we can assign numerical values to each rating and perform basic statistical calculations. The table shows the frequency of each of the three rating options for all criteria, as well as the weighted relevance score S.

We have defined the weighted relevance score (S) for each information quality criterion as the measure of its relevance. The weighted score $S_i$ for an information quality criterion *i*, given N respondent answers, is calculated as follows:

$$S_i = 100 \frac{\sum_1^N Answer_{Relevant} + \frac{1}{2}\sum_1^N Answer_{SomewhatRelevant}}{N}$$

The column 'Relevant' in Table 4 shows numbers in bold when the frequency is less than 50% of the total number of answers.

| Information Criteria | Relevant | Somewhat Relevant | Not Relevant | S |
|---|---|---|---|---|
| Accuracy (The extent to which information corresponds to the true value) | 35 | | | 100 |
| Precision (The extent to which different measurements of the information are close to each other) | 19 | 15 | 1 | 76 |
| Reputation (The extent to which information is highly regarded in terms of its source or content) | 18 | 11 | 6 | 67 |
| Objectivity (The extent to which information is unbiased, unprejudiced, and impartial) | 21 | 13 | 1 | 79 |
| Traceability (The extent to which information's source can be found) | 25 | 10 | | 86 |
| Appropriate Amount (The extent to which the volume of information is appropriate for the task at hand) | **17** | 14 | 4 | 69 |
| Ease of Use (The extent to which information is easy to manipulate and apply to different tasks) | **13** | 17 | 5 | 61 |
| Effectiveness (The extent to which information is able to quickly meet the information needs for the task at hand) | 20 | 12 | 3 | 74 |
| Completeness (The extent to which information is not missing and is of sufficient depth and breadth for the task at hand) | 27 | 8 | | 89 |
| Relevancy (The extent to which information is applicable and helpful for the task at hand) | 24 | 10 | 1 | 83 |
| Currency (The extent to which information is sufficiently up to date for the task at hand) | 25 | 7 | 3 | 81 |
| Timeliness (The extent to which the age of the data is appropriated for the task at hand) | 22 | 12 | 1 | 80 |
| Interpretability (The extent to which information is in appropriate languages, symbols and units, and the definitions are clear) | 25 | 7 | 3 | 81 |
| Consistency (The extent to which information is presented in the same format) | 21 | 10 | 4 | 74 |
| Conciseness (The extent to which information is compactly represented) | **12** | 12 | 11 | 51 |
| Understandability (The extent to which information is easily comprehended) | 21 | 12 | 2 | 77 |
| Accessibility (The extent to which access to information is restricted appropriately to authorized parties) | 19 | 9 | 7 | 67 |
| Availability (The extent to which information is available when required, or easily and quickly retrievable) | 20 | 9 | 6 | 70 |

Table 8–- Analysis of information quality criteria based on the answers to closed question Q2

We used the weighted relevance score S as the main decision criterion for deciding whether an information quality criterion can be considered validated, provisionally validated, or rejected by the expert panel. We consider the criterion validated if S >= 70, provisionally validated if 70 >S >=60 and rejected if S < 60. Provisionally validated criteria will be individually discussed for their disposition.

Using the thresholds defined above, the following criteria are validated: 1. Accuracy (100), 2. Precision (76), 4. Objectivity (79), 5. Traceability (86), 8. Effectiveness (74), 9. Completeness (89), 10. Relevancy (83), 11. Currency (81), 12. Timeliness (80), 13. Interpretability (81), 14. Consistency (74), 16. Understandability (77), and 18. Availability (70).

Based on the same thresholds, the following criterion is rejected: 15. Conciseness (51).

The remaining criteria are provisionally validated. We discussed whether to maintain or reject them based on the considerations documented in Table 9:

| Information Criteria | Relevant | Somewhat Relevant | Not Relevant | S |
|---|---|---|---|---|
| Reputation (The extent to which information is highly regarded in terms of its source or content) | 18 | 11 | 6 | 67 |
| Although below the threshold for validation, we decided to maintain the criterion in the model for the following reasons: The number of 'relevant' ratings exceeded 50%, as well as the number of 'somewhat relevant' ratings for the remaining experts. The S score was only marginally below the threshold, and the fact that the current explanation and labelling is probably not ideal (i.e., 0 mentions after step 1 but 17 mentions after step 2 – see Table 5) made us decide to maintain it in the model but to improve the description and labelling of this criterion. | | | | |
| Appropriate Amount (The extent to which the volume of information is appropriate for the task at hand) | **17** | 14 | 4 | 69 |
| Although below the threshold for validation, we decided to maintain the criterion in the model for the following reasons:<br>The S score was only very marginally under our threshold, the number of 'relevant' scores was only marginally under 50% whereas the number of 'somewhat relevant' scores was very high for the remaining experts that did not rate the criterion as 'Relevant'. | | | | |
| Ease of Use (The extent to which information is easy to manipulate and apply to different tasks) | **13** | 17 | 5 | 61 |
| This criterion scores well below the threshold, and because less than half of the respondents see this as a relevant criterion, we decided to reject this criterion and remove it from the model. | | | | |
| Accessibility (The extent to which access to information is restricted appropriately to authorized parties) | **19** | 9 | 7 | 67 |
| Although below the threshold for validation, we decided to maintain the criterion in the model for the following reasons: the number of 'relevant' ratings exceeded 50%, as well as the number of 'somewhat relevant' scores for the remaining experts. The S score was only marginally below the threshold. | | | | |

Table 9 - Provisionally Validated Criteria and Justified Decisions

Thirteen experts (39%) provided comments on question Q2. These comments are listed in full in Appendix 6. They can be summarized as follows:

- Several respondents (5 out of 13) commented that the relevance of criteria will depend on context or type of information. This confirms our assumption and the answers to Q3 will be investigated to find out if our assumption is further supported.
- Several remarks were made on unclear definitions or overlap between criteria: Timeliness and Currency (3 out of 13 respondents), Completeness and Appropriate Amount, and Availability and Accessibility (2 out of 13 respondents). We consider reviewing these definitions and updating them where appropriate during the final artefact design.
- One expert related the availability of measurement information and the timing of such measurement on the proposed criterion to the relevance of the criterion – we acknowledge the remark but find it not relevant to the existence and potential relevance of the information quality criteria.
- A few comments questioned the validity of some criteria (i.e., Objectivity, Accessibility, Availability, Reputation), but that was also reflected in the 'Not Relevant' score for those criteria.
- One expert suggested also some missing criteria (i.e., Atomicity, Durability) which will be further discussed concerning questions Q4 and Q5.

Finally, we compared the results of open question Q1 with the results of closed question Q2. For Q2 we took the frequency of 'Relevant' answers as the basis for this comparison because we think that only relevant criteria would be spontaneously mentioned as answers to Q1. The result of this comparison is shown in Figure 2.

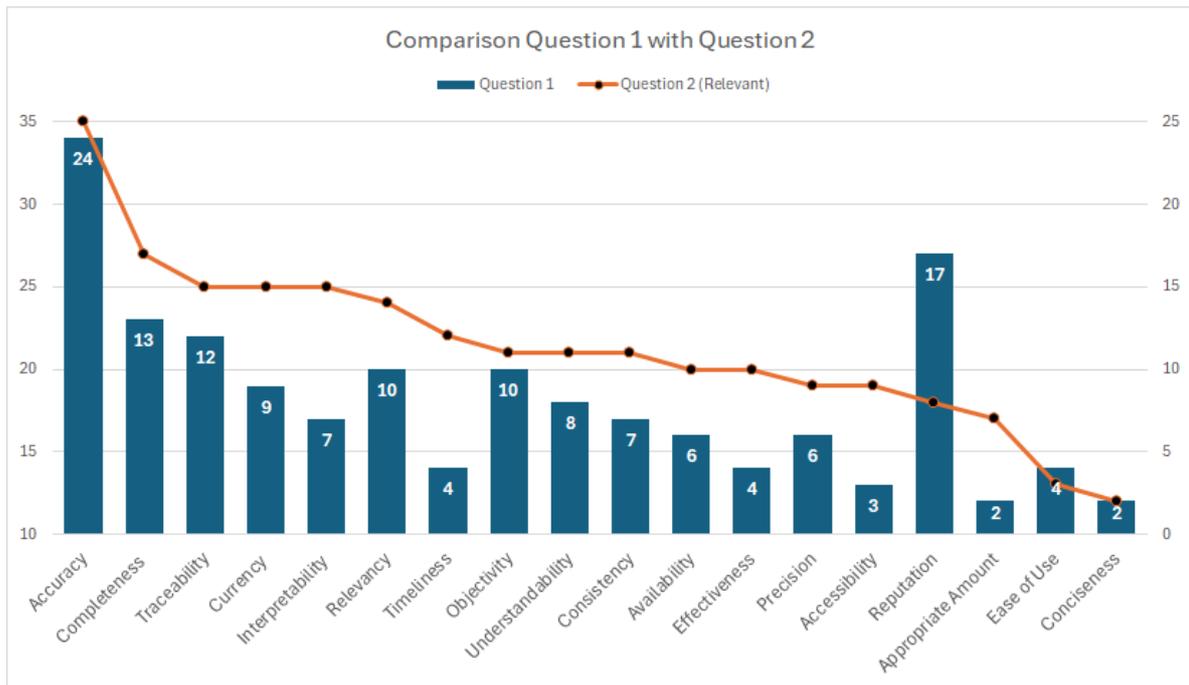

Figure 2–- Mapping between the Q1 and Q2 answer sets

The Pearson Correlation between the two data sets is 0.74, which indicates a strong positive linear relationship between the two answer sets. However, we visually observe that Reputation is a clear outlier (if we remove the criterion, the Pearson Correlation is 0.90), being very frequently mentioned in response to Q1 but receiving a relatively low score on relevance in the answers to Q2. A possible explanation for this phenomenon is that the definition of this information quality attribute in the second iteration design of the reference model distracts from its intended meaning. As already indicated in Table 9, we will review the label and definition of this criterion in the final artefact design.

*Question 3: Can you think of any examples where the importance of quality criteria depends on the purpose the information is serving? (N=29)*

The purpose of this question was to probe for the presence of contingency factors, i.e., whether the importance of an information quality criterion varies between different information items or not. The results of this question can be found in Appendix 7 and can be summarized as follows:

- The importance of information quality criteria is not absolute but rather seems to depend on the nature of the information itself. In other words, there are contingency factors for the relevance of information criteria that need to be identified.
- The high number of examples that were given exceeded our expectations (N=71 by 29 respondents), demonstrating the relevance of the question, as well as the engagement of the expert panel.
- A few answers were too short to include in the analysis, e.g., when just mentioning a criterion or an information item.
- Many examples were structured as "<criterion X> is more/less important for <information item Y> in <context A>". Other examples were structured as "<criterion X> is more/less important for <information item Y> than for <information item Z> in <context A>".
  - Many of the criteria of the second iteration design of the reference model were used in the examples, with varying frequencies (Figure 3).

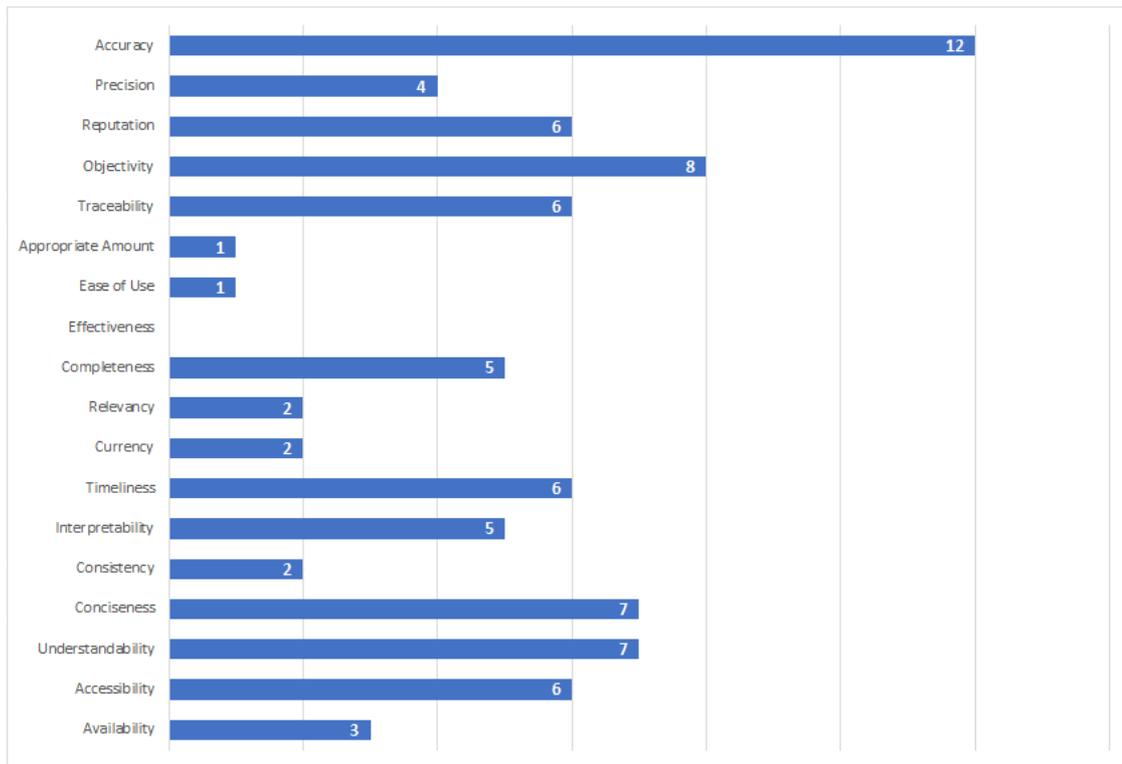

Figure 3 – Frequency table of the Information Quality Criteria in examples on contingencies (Q3)

We cannot use these data to draw any firm conclusions about the (in)consistency of these data with the results of questions Q1 and Q2, because a mention of a criterion here can both indicate a positive (more relevant) or a negative (less relevant) relationship. However, it is interesting to observe the confirmation of our assumption that the relevance of information quality criteria may vary between information items depending on the purpose of using them.

Given that our panel was largely composed of experts in IT Governance and related areas, this raises the question of which contingency factors in IT Governance systems make information quality criteria relevant for assessing the quality of certain IT Governance information items, which is a question we take up in future research.

*Q4. Would you see any additional criteria that you think are relevant for defining and assessing information quality?*

*Q5. Which additional criteria according to you are relevant for information quality?*

The large majority of experts (29 out of 35, or 83%) did not identify any additional information quality criteria. The few suggestions that were made were not retained for inclusion in the final artefact design because they were covered by already existing attributes. The detailed analysis can be found in Appendix 8.

*Third Iteration or Final Artefact Design*

From the results of the expert validation, we conclude that the information quality criteria in the second iteration design of the reference model are largely validated, but with some criteria to be removed, and some modifications in the definitions and descriptions required. Table 10 describes the changes made to arrive at the final design of the IQRM.

| Information Quality Criteria | Disposition and Comment |
|---|---|
| Source Reliability (The extent to which information is highly regarded in terms of its source or content) | Relabeled for clarity – original label was 'reputation' |
| Traceability (The extent to which information's source or sources can be found) | Reworded for clarity – original text was 'The extent to which information's source can be found' |
| Ease of Use (The extent to which information is easy to manipulate and apply to different tasks) | Rejected because of lack of relevance. |
| Effectiveness (The extent to which information is able to meet the information needs for the task at hand) | Relabeled for clarity – original label was 'efficiency' |
| Timeliness (The extent to which the moment in time the data is available for the task at hand is appropriate) | Reworded for clarity – original text was 'The extent to which the age of the data is appropriated for the task at hand' |
| Interpretability (The extent to which information is in appropriate languages, symbols and units, and the definitions are clear for both human and machine users) | Reworded for clarity – original text was 'The extent to which information is in appropriate languages, symbols and units, and the definitions are clear' |
| Conciseness (The extent to which information is compactly represented) | Rejected because of lack of relevance |

Table 10 - Conclusion from Information Quality Criteria Validation

Based on the conclusions above, we finalized the IQRM as shown in Table 11. Because of the removal of Ease of Use and Conciseness, the design of the IQRM could be simplified by also removing the quality attributes of which these criteria were sub-attributes, respectively Usability and Clarity.

| Information Quality Reference Model – Final DESIGN | | | |
|---|---|---|---|
| **Quality Dimension** | **Quality Sub-Dimension** | **Quality Attributes and Sub-Attributes** | |
| Intrinsic<br>The extent to which data values are in conformance with the actual or true values | Correctness<br>The extent to which the data is accurate and precise | Accuracy | The extent to which information corresponds to the true value |
| | | Precision | The extent to which different measurements of the information are close to each other |
| | | Source Reliability | The extent to which information is highly regarded in terms of its source or content |
| | Reliability<br>The extent to which information is believable, reputed highly and objective | Objectivity | The extent to which information is unbiased, unprejudiced and impartial |
| | | Traceability | The extent to which information's source or sources can be found |
| Contextual<br>The extent to which information is applicable to the task of the information user | Adequacy<br>The extent to which information is sufficient for the task at hand | Appropriate Amount | The extent to which the volume of information is appropriate for the task at hand |
| | | Effectiveness | The extent to which information is able to meet the information needs for the task at hand |
| | | Completeness | The extent to which information is not missing and is of sufficient depth and breadth for the task at hand |
| | | Relevancy | The extent to which information is applicable and helpful for the task at hand. |
| | | Time<br>The extent to which information is timely and current | Currency: The extent to which information is sufficiently up to date for the task at hand |
| | | | Timeliness: The extent to which the age of the data is appropriated for the task at hand |
| | Representation<br>The extent to which information is presented in an intelligible and clear manner | Interpretability | The extent to which information is in appropriate languages, symbols and units, and the definitions are clear for both human and machine users |
| | | Consistency | The extent to which information is presented in the same format |
| | | Understandability | The extent to which information is easily comprehended |
| Security<br>The extent to which information is available and accessible | Accessibility | | The extent to which access to information is restricted appropriately to authorized parties |
| | Availability | | The extent to which information is available when required, or easily and quickly retrievable |

Table 11 – Final IQRM Artefact Design

**Discussion and Conclusion**

*Outcomes*

The results allow answering our research questions RQ1 and RQ2 as follows.

For RQ1 (*Which attributes of information constitute a practical and comprehensive reference model for information quality in the context of IT Governance?*) the answer is embedded in our design artefact, the IQRM. This reference model for IT Governance information quality contains 16 information quality criteria that are organized as (sub-) attributes of information quality along two sub-dimensions of intrinsic information quality (i.e., correctness and reliability), two-subdimensions of contextual information quality (i.e., adequacy and representation), and a security dimension. The IQRM has been iteratively developed and evaluated, using a Design Science approach, based on information quality models from literature and the domain knowledge of a panel of 35 IT Governance experts.

For RQ2 (*To what extent is the importance of these information quality attributes dependent on the purpose the information serves in an IT Governance system*) we found that, according to the IT Governance experts of our panel, not all the information quality criteria we identified are equally important for all information items. A more precise answer to RQ2 will depend on contingency factors to be identified in further research.

Furthermore, as a solution designed for a problem experienced in practice, we observe that the IQRM contains an elaborate yet manageable number of criteria for information quality. The number of criteria for information quality is sufficiently high to allow it to be comprehensive and granular. At the same time, it is not too high not to overburden users of the model.

*Comparison with Related Work in the Practitioner Community*

Our initial research started in 2012 during the development of the COBIT 5 Framework. Since then, other frameworks or standards for ensuring information quality have been published in the grey literature. Some of these works have a similar focus to ours on defining information quality in terms of dimensions and attributes that could serve as information quality criteria for performance assessment, although none of the reviewed frameworks is specifically intended to be used in the IT Governance context. For the sake of completeness, we compare the IQRM with two of these frameworks: Myer's Conformed Dimensions of Data Quality (Myers, 2015) (Table 12) and the DMBOK2 data quality dimensions (DAMA, 2017) that were already listed in Table 2.

| Conformed Dimension | Definition |
|---|---|
| Completeness | Completeness measures the degree of population of data values in a data set. |
| Accuracy | Accuracy measures the degree to which data factually represents its associated real-world object, event, |
| Consistency | Consistency measures whether data is equivalent across systems or location of storage. |
| Validity | Validity measures whether a value conforms to a preset standard. |
| Timeliness | Timeliness is a measure of time between when data is expected versus made available. |
| Currency | Currency measures how quickly data reflects the real-world concept that it represents. |
| Integrity | Integrity measures the structural or relational quality of data sets. |
| Accessibility | Accessibility measures how easy it is to acquire data when needed, how long it is retained, and how access is controlled. |
| Precision | Precision is the measurement or classification detail used in specifying an attribute's domain. |
| Lineage | Lineage measures whether factual documentation exists about where data came from, how it was transformed, where it went and end-to-end graphical illustration. |
| Representation | Representation measures ease of understanding data, consistency of presentation, appropriate media choice, and availability of documentation (metadata). |

Table 12 - Myer's Conformed Dimensions of Data Quality (DQMatters, 2019-2024)

Table 13 shows a comparison of the three models.

| IQRM | Conformed Dimensions of Data Quality | DMBOK2 |
|---|---|---|
| ACCURACY | ACCURACY | ACCURACY / INTEGRITY |
| PRECISION | PRECISION | |
| SOURCE RELIABILITY | LINEAGE | |
| OBJECTIVITY | | |
| TRACEABILITY | LINEAGE | |
| APPROPRIATE AMOUNT | | |
| EFFECTIVENESS | | |
| COMPLETENESS | COMPLETENESS | COMPLETENESS / INTEGRITY |
| RELEVANCY | | |
| CURRENCY | CURRENCY | |
| TIMELINESS | TIMELINESS | TIMELINESS |
| INTERPRETABILITY | REPRESENTATION | |
| CONSISTENCY | CONSISTENCY | CONSISTENCY/INTEGRITY |
| UNDERSTANDABILITY | REPRESENTATION | |
| ACCESSIBILITY | ACCESSIBILITY | |
| AVAILABILITY | | |
| | INTEGRITY | |
| | VALIDITY | VALIDITY |
| | | UNIQUENESS |
| | | REASONABILITY |

Table 13 – Comparison of the IQRM with Myers Conformed Dimensions of Data Quality and DMBOK2

When comparing the information quality criteria of the three models, we observe similarities and differences. Focusing on the differences we would argue that the dimensions retained in Myers' model and in DMBOK2 are very much related to data in the strict sense, i.e. the more atomic basic pieces of (typically) structured information as we store them in databases. This applies, for instance, to Validity, Uniqueness and Reasonability. We believe those dimensions or attributes are less relevant for information. Given our scope of IT Governance and the information flows or items that are required to support an IT Governance system, we must deal almost exclusively with 'data in use' as opposed to 'data residing in systems' or 'data being processed' *(*Madnick et al., 2009). From this perspective, the quality attributes that we identified and that are not part of the other models, i.e., Relevancy, Effectiveness, Objectivity, Appropriate Amount, and Availability, make sense.

We conclude from the comparison that the models, especially the IQRM and the Conformed Dimensions of Data Quality, align quite well, and that the observed differences do not indicate either model is inadequate but rather are caused by their slightly different focus on information versus data.

***Contribution and Implications***

The main contribution of this paper is the Information Quality Reference Model (IQRM), which is an empirically validated reference model for IT Governance information quality that was developed based on information quality models found in the literature. In the first instance, this model will help to assess the quality of information in an IT governance context, i.e., assessing the quality of the information items we need as one of the components of the IT Governance system of an organization (ISACA, 2018a). In addition, this model has the potential to form the basis for developing a capability model for said information items. This will allow practitioners to make a more thorough assessment of the quality of the information used in their IT Governance practices and decisions, as well as a more complete assessment of their IT Governance systems. Researchers can use the IQRM to investigate research questions on the quality of information used in IT Governance systems and their impact on the overall performance of an IT Governance system. The IQRM can also be used to specify quality requirements for information or to assess actual quality to improve quality where necessary.

As a final observation – because we did not specify any specific IT Governance information items during the iterative development and evaluation of the IQRM – we believe this model is a good candidate for a generic information quality model. As such it could become a useful model for the assessment and management of quality of information at large.

*Limitations*

We have researched quite comprehensively the existing literature on information quality focusing on the meaning of the concept of information quality and how to decompose this concept along dimensions and attributes. However, the literature review was not a systematic literature review so there remains the likelihood that we missed some important sources.

We believe that we have proposed a model that is universally useable within an IT Governance context, as our expert panel included primarily experts in the IT Governance domain. The validation by our expert panel could be further strengthened by either enlarging the panel or by conducting case studies in information-intensive organizations.

During our research, we had to make multiple interpretations when looking at the naming and description of information quality attributes and when interpreting and coding the expert panel's responses. Instead of the manual interpretation and internal team validation we have performed, we could also have used automated tools or AI tools for the qualitative analysis.

*Further Research Opportunities*

Summarizing some of the remarks made above, further research is possible to better understand and validate the proposed artefact, e.g., conduct further empirical research with a wider audience or through case studies about the acceptance, relevance, and completeness of the model. Some areas for potential further research include:

- The scope of the model could be widened from an IT Governance context to a context of information quality at large, requiring differently composed expert panels and/or additional research methods.

- Regarding the model itself, further research could be conducted on its potential for wider adoption, e.g. using a TAM (Davis, 1993) or UTAUT-like (Venkatesh et al, 2003) approach.
- When thinking about the operationalization of the IQRM, research on metrics for each of the quality attributes could also be very useful.
- Research on the potential relationship between the quality of data management or data quality processes and the actual quality of information – using the IQRM – could lead to interesting insights.
- We have suggested the opportunity that this reference model for information quality could serve as the foundation for a capability model for IT Governance information, and this opportunity we take up as future research ourselves.
- In this paper we have also identified the probable existence of contingency factors for the importance of information quality criteria, i.e., a given quality criterion will not be equally important for different information items. The nature and extent of these contingency factors need to be further researched, which is also on our research agenda.

## Appendix 1 – Expert Panel Details

This appendix contains more detailed information on the composition of the expert panel and the following information on individual participants: age group, experience, function & expertise area, and nationality.

| # | Age Group[1] | # Years Experience | Function & Expertise Area | Name | Nationality |
|---|---|---|---|---|---|
| 1 | 30-40 | 16 | Senior manager in large consulting company Experienced consultant - IT strategist and enterprise architect | Known | Germany |
| 2 | 30-40 | 13 | Director Data & Analytics, Data Strategy & Governance in large consulting company – specialist data governance | Known | Belgium |
| 3 | 25-30 | 6 | Senior technology consultant IT Governance, IT Strategy. | Known | Belgium |
| 4 | | | Anonymous | | Argentina |
| 5 | 40-50 | 21 | Enterprise & Solution architect at consulting firm – expert in enterprise architecture & solution architecture | Known | Belgium |
| 6 | 40-50 | 25 | Experienced Consultant – fields of expertise include Enterprise architecture, compliance and information security | Known | Belgium |
| 7 | 30-40 | 20 | CISO (Chief Information Security Officer) at division level of large multinational manufacturing company. – experienced professional in information security, architecture and service delivery. | Known | Belgium |
| 8 | 30-40 | 16 | Solution architect with extensive experience in software architecture and systems delivery | Known | Belgium |
| 9 | | | Anonymous | | Belgium |
| 10 | 30-40 | 12 | Information Security Specialist, IT and data governance & audit | Known | Belgium |
| 11 | 50+ | 28 | Full professor at top 100 university Academic research on Information systems, conceptual modelling, architecture, governance | Known | Belgium |
| 12 | 25-30 | 3 | Consultant at large consulting firm – active in field of IT Governance | Known | Belgium |
| 13 | 50+ | 27 | IT Governance consultant – expert in IT Governance, Cybersecurity, IT Management | Known | USA |
| 14 | 30-40 | 7 | Assistant Professor in Information Systems strategic management and IT Governance. | Known | Belgium |
| 15 | 50+ | 32 | Independent Consultant in Strategy & Compliance in environmental recycling systems and circular economy | Known | Belgium |

---

[1] We have classified age into the following groups: 25-30 years old; 30-40 years old, 40-50 years old, 50+ years old.

| # | Age Group[1] | # Years Experience | Function & Expertise Area | Name | Nationality |
|---|---|---|---|---|---|
| 16 | 50+ | 31 | Partner in international consulting firm with background in cybersecurity, IT Governance, IT strategy. Entrepreneur. | Known | Belgium |
| 17 | 40-50 | 14 | Legal specialist in environmental international law | Known | Chile |
| 18 | 50+ | 24 | Assistant Professor of Information Systems | Known | Portugal |
| 19 | 50+ | 37 | Experienced languages and literature teacher | Known | Belgium |
| 20 | 50+ | 35 | Experienced technology consultant, now CIO in healthcare; extensive background in IT auditing, IT service management, IT governance. | Known | Belgium |
| 21 | 30-40 | 10 | Digital transformation manager and IT program manager in major construction group. | Known | Belgium |
| 22 | 30-40 | 16 | Manager in major chemical plant in data protection, compliance, security, digital innovation. | Known | Belgium |
| 23 | | | Anonymous | | Belgium |
| 24 | | | Anonymous | | Belgium |
| 25 | 30-40 | 8 | Sales and Business Development in telco industry, pursuing Master in Enterprise architecture | Known | Lithuania |
| 26 | 50+ | 34 | Vice president IT and smart cities at regional government agency | Known | Belgium |
| 27 | 50+ | 25 | Head of IT at government agency | Known | Belgium |
| 28 | 40-50 | 18 | Head of Sector (IT Governance) at European Commission's IT Directorate | Known | Poland |
| 29 | 30-40 | 11 | Experienced consultant in information security, privacy and risk management | Known | Belgium |
| 30 | 50+ | 28 | Vice President Content Development of international professional association – earlier experience in information publishing industry | Known | USA |
| 31 | 30-40 | 9 | Consultant Business Intelligence and business transformation projects | Known | Germany |
| 32 | 50+ | 35 | Professor in IT management, Cybersecurity and IT Governance at top-100 university | Known | Belgium |
| 33 | 50+ | 29 | Executive management consulting in IT Security, IT Governance, IT Risk Management | Known | South Africa |
| 34 | 30-40 | 13 | Experienced IT Consultant enterprise architecture, lecturer, entrepreneur | Known | Belgium |
| 35 | 50+ | 29 | Director of technology & security assurance, Chief security officer - experienced senior manager in information security, information assurance | Known | Australia |

The following table contains a description of the expert panel members and relevant information on their expertise supporting their presence on the panel. All expert panel's names are known to the author but are left out of this table for privacy reasons, except for four panel members who choose to remain anonymous.

## Age distribution

| | | |
|---|---|---|
| 25-30 | 2 | 6% |
| 30-40 | 12 | 35% |
| 40-50 | 4 | 11% |
| 50+ | 13 | 37% |
| Unknown | 4 | 11% |

## Relevant Experience distribution

| | | |
|---|---|---|
| <10 | 5 | 14% |
| Between 10 and 15 | 6 | 17% |
| Between 15 and 20 | 4 | 11% |
| Between 20 and 30 | 10 | 29% |
| >30 | 6 | 17% |
| Unknown | 4 | 11% |

## International distribution

| | | |
|---|---|---|
| Belgium | 24 | 68% |
| USA | 2 | 6% |
| Germany | 2 | 6% |
| Other | 7 | 20% |

# Appendix 2 – Questionnaire for the Expert Panel Survey on Information Quality Criteria validation

The following questionnaire was used with the expert panel for the validation of the proposed Information Reference Model:

*Q1. When you consider 'information quality', what are the criteria you would use to decide whether the information has quality or not? (Information can be any type of information or data)*

*Q2. To what extent do you think each of the following 18 criteria would be relevant when assessing the quality of information, regardless of the purpose of the information?*

| Information Quality Criteria | Relevant | Somewhat Relevant | Not Relevant |
|---|---|---|---|
| Accuracy (The extent to which information is semantically correct) | ☐ | ☐ | ☐ |
| Precision (The extent to which information is syntactically correct) | ☐ | ☐ | ☐ |
| Reputation (The extent to which information is highly regarded in terms of its source or content) | ☐ | ☐ | ☐ |
| Objectivity (The extent to which information is unbiased, unprejudiced, and impartial) | ☐ | ☐ | ☐ |
| Traceability (The extent to which information's source can be found) | ☐ | ☐ | ☐ |
| Appropriate Amount (The extent to which the volume of information is appropriate for the task at hand) | ☐ | ☐ | ☐ |
| Ease of Use (The extent to which information is easy to manipulate and apply to different tasks) | ☐ | ☐ | ☐ |
| Efficiency (The extent to which information is able to quickly meet the information needs for the task at hand) | ☐ | ☐ | ☐ |
| Completeness (The extent to which information is not missing and is of sufficient depth and breadth for the task at hand) | ☐ | ☐ | ☐ |
| Relevancy (The extent to which information is applicable and helpful for the task at hand) | ☐ | ☐ | ☐ |
| Currency (The extent to which information is sufficiently up to date for the task at hand) | ☐ | ☐ | ☐ |
| Timeliness (The extent to which the age of the data is appropriated for the task at hand) | ☐ | ☐ | ☐ |
| Interpretability (The extent to which information is in appropriate languages, symbols and units, and the definitions are clear) | ☐ | ☐ | ☐ |
| Consistency (The extent to which information is presented in the same format) | ☐ | ☐ | ☐ |
| Conciseness (The extent to which information is compactly represented) | ☐ | ☐ | ☐ |
| Understandability (The extent to which information is easily comprehended) | ☐ | ☐ | ☐ |
| Accessibility (The extent to which access to information is restricted appropriately to authorized parties) | ☐ | ☐ | ☐ |
| Availability (The extent to which information is available when required, or easily and quickly retrievable) | ☐ | ☐ | ☐ |

Please put here any comment on the criteria above, e.g. if you think there are duplicates.

*Q3. Can you think of any examples where the importance of quality criteria depends on the purpose the information is serving? <u>Example</u>: timeliness will be much more important for a financial transaction compared to a yearly management plan.*

*Q4. Would you see any additional criteria that you think are relevant for defining and assessing information quality?*

*Q5. Which additional criteria according to you are relevant for information quality?*

*Q6. Are you interested in participating in a follow-on survey on information quality and maturity models?*

*Q7. Thank you for your interest in the follow-on survey. Which e-mail address can we use to contact you (only to send a link for the follow-on survey, no other contacts will be made)*

# Appendix 3 – Survey on Information Quality Criteria validation – Detailed answers Question 1

The following was mentioned by the respondents that completed this question[2] [3]

(N=35):

| | |
|---|---|
| 1 | Fitness for use, so depending on the context and the intended usage, the criteria can be different and threshold for being considered as qualitative might vary. Dimensions to be potentially be taken into account are: completeness, accuracy, timeliness, validity, consistency, conformance, relevance, interpretability, .. |
| 2 | correct, complete, relevant, integer, timely (= data covers the period I'm interested in), available |
| 3 | Accuracy, Source trustworthiness, Relevancy, Completeness, Currentness |
| 4 | The meaning (what the information represents) is clear, The context is clear, The Reliability of the information is high ( know where it comes from and how it has been generated), The Information is precise |
| 5 | Accuracy, recency, reliability, relevance, availability/accessibility, objectivity, interpretability/understandability, coherence, completeness |
| 6 | Complete, correct and minimal explaining what, why and how in a story telling manner |
| 7 | Relevant, with just the right amount of detail |
| 8 | Accurate and time bound data, most of all fit for purpose (how good is the data for our needs). |
| 9 | correct, understandable, not confusing, good argumentation |
| 10 | Accuracy of data, Relevance of data/ relevance of information, Format/ presentation, Ease of processing, Transparency of sources used, Ease of information consumption |
| 11 | Was the information useful? |
| 12 | Completeness (missing records/chapters), timing (recent/real-time vs. outdated information), structure, correctness (typos), information origin/source, realistic vs. fictional information |
| 13 | Does it make sense in the context that I want to use it? |
| 14 | Accurate, objective, relevant, complete, current, understandable |
| 15 | Availability, usability, reliability |
| 16 | source is reliable, fact based analysis, verifiable |
| 17 | accurate (completeness, existence, ..), relevant, secure (incl availability, restricted access) |
| 18 | Reliability of the sources, completeness, coherence. |
| 19 | availability and reliability |
| 20 | reliable source/author, recent date, frequency of occurrence, frequency of consultation/use, probability of correctness |
| 21 | Reliability & accuracy |
| 22 | understandable (structure, representation), completeness, accuracy, |
| 23 | Completeness, Correctness, depending on the type: up-to-date or not? (Old information can still be relevant for evolutions etc.), Trustworthiness. (Are we sure the information is untampered with? Do we know the source?) |
| 24 | Information should be accurate, up-to-date, reliable, trustworthy, protected against undesirable modification, available when needed, limited to what is needed, easily accessible and secure. |
| 25 | The relevance of the data, Objective results, different sources compared |

---

[2] Answers are represented as provided by the respondents, with typos and grammatical errors corrected and translations (using google translate) into English when answers were given in a different language.

[3] The numbers in front of the answers do not correspond to the numbers in front of the expert panel members' details in Appendix 1.

| 26 | Mainly the source of the information |
|---|---|
| 27 | That it's accurate across various platforms for all users (example: sales figures: same revenues, same invoiced values) it is somehow cleaned - no raw data for wrong interpretation. |
| 28 | Traceerbaarheid - ken ik de bron en wie /wat is de bron - zegt dus iets over de betrouwbaarheid van de informatie/data. Is het bewerkte informatie zo ja in welke zin. (*Traceability - do I know the source and who/what is the source - therefore says something about the reliability of the information/data. Is it edited information, if so in what sense?*) |
| 29 | When the information is well definied, the semantic meaning is clear to everybody and the information is in accordance with standards and definied structures. Metadata is present. There is some kind of version control in place. |
| 30 | Being up to date and accurate |
| 31 | From a managerial point of view it has to be authoritative (central versions of truth need to be clear across multiple information repositories), concise ('just enough', aligned with stakeholder needs) and embedded in (automated) processes in such a way (from creation to ETL operations to reporting) that the stakeholders are incentivized to keep the information accurate and up-to-date. From a technical point of view it has to be created transactional (in accordance to the ACID principles) and linked with other information to ensure consistency. The level of information detail should also be consistent, either agreed across all stakeholders or tailored on a per stakeholder level. |
| 32 | Is the data accurate? Is the data formatted correctly (e.g., is it consistent)? Is the data complete?  In short, is the data fit for the purpose for which I need? |
| 33 | The information is coherent, true and can be used without explanation |
| 34 | effective, efficient, reliable, timely, |
| 35 | Accuracy, Completeness, Authenticity, Auditable, Relevant, Traceable |

# Appendix 4 – Detailed Analysis Question 1 – Analysis Step 2

This appendix contains step 2 of the analysis of the answers to question 1 of the survey. For each answer received the table contains a "i" if the quality criterion (as listed on the first row of the table) appeared literally in the answer (as identified in step 1 of the analysis), and a "ii" if we assessed that the answer contained a synonym for the quality criterion. The parts of the answer matched literally are in strikethrough font, and the parts of the answers that have been assessed as synonyms are printed in red font. The remaining parts of the answers – if any – are further analyzed and discussed in step 3.

For readability reasons, the table is split in three parts, each containing all answers and a subset of the information quality criteria.

|  | Accuracy | Precision | Reputation | Objectivity | Traceability | Appropriate Amount | Ease of Use |
|---|---|---|---|---|---|---|---|
| 1) Dimensions to be potentially be taken into account are: ~~completeness, accuracy, timeliness~~, validity, ~~consistency~~, conformance, ~~relevance, interpretability~~, .. | ii | ii |  |  |  |  |  |
| 2) correct, ~~complete, relevant~~, integer, ~~timely (= data covers the period I'm interested in), available~~ | ii | ii |  | ii |  |  |  |
| 3) ~~Accuracy~~, Source trustworthiness, ~~Relevancy, Completeness, Currentness~~ | i |  | ii |  |  |  |  |
| 4) The meaning ( what the information represents) is clear, The context is clear, The Realibilty of the information is high ( know where it comes from and how it has been generated), ~~The Information is precise~~ |  | i | ii |  |  |  |  |
| 5) ~~Accuracy~~, recency, reliability, ~~relevance, availability/accessibility, objectivity, interpretability/understandability~~, coherence, ~~completeness~~ | i |  | ii | i | ii |  |  |
| 6) ~~Complete~~, correct and minimal explaining what, why and how in a story telling manner | ii | ii |  |  |  |  |  |
| 7) ~~Relevant~~, with just the right amount of detail |  |  |  |  |  | ii |  |
| 8) ~~Accurate~~ and time bound data, most of all fit for purpose (how good is the data for our needs). | i |  |  |  |  |  | ii |

| | Accuracy | Precision | Reputation | Objectivity | Traceability | Appropriate Amount | Ease of Use |
|---|---|---|---|---|---|---|---|
| 9) correct, ~~understandable~~, not confusing, good argumentation | ii | ii | | ii | | | |
| 10) ~~Accuracy of data~~, ~~Relevance of data/ relevance of information~~, Format/ presentation, Ease of processing, Transparency of sources used, ~~Ease of information consumption~~ | i | | ii | | | | i |
| 11) Was the information useful? | | | | | | | ii |
| 12) ~~Completeness (missing records/chapters)~~, ~~timing (recent/real-time vs. outdated information)~~, structure, correctness (typos), information origin/source, realistic vs. fictional information | ii | | ii | | | | |
| 13) Does it make sense in the context that I want to use it? | | | | | | | |
| 14) ~~Accurate, objective, relevant, complete, current, understandable~~ | i | | | i | | | |
| 15) ~~Availability~~, usability, reliability | | | ii | ii | ii | | ii |
| 16) source is reliable, fact based analysis, verifiable | | | ii | | ii | | |
| 17) ~~accurate (completeness, existence, ..), relevant, secure (incl availability, restricted access)~~ | i | | | | | | |
| 18) Reliability of the sources, ~~completeness~~, coherence. | | | ii | | | | |
| 19) ~~availability~~ and reliability | | | ii | ii | ii | | |
| 20) reliable source/author, ~~recent date~~, frequency of occurrence, frequency of consultation/use, probability of correctness | ii | | ii | | ii | | |
| 21) Reliability & ~~accuracy~~ | i | | ii | ii | ii | | |
| 22) ~~understandable (structure, representation), completeness, accuracy,~~ | i | | | | | | |
| 23) ~~Completeness~~, Correctness, ~~depending on the type: up to date or not? (Old information can still be relevant for evolutions etc.)~~, Trustworthiness. (Are we sure the information is untampered with? Do we know the source?) | ii | ii | ii | | ii | | |
| 24) Information should be ~~accurate, up to date~~, reliable, trustworthy, protected agains undesirable modification, ~~available when needed~~, limited to what is needed, ~~easily accessible~~ and secure. | i | | ii | ii | ii | ii | |
| 25) ~~The relevance of the data, Objective results~~, different sources compared | | | ii | i | ii | | |

| | Accuracy | Precision | Reputation | Objectivity | Traceability | Appropriate Amount | Ease of Use |
|---|---|---|---|---|---|---|---|
| 26) Mainly the source of the information | | | ii | | | | |
| 27) ~~That it's accurate across various platforms for all users (example: sales figures: same revenues, same invoiced values)~~ it is somehow cleaned - no raw data for wrong interpretation. | i | | | | | | |
| 28) ~~Traceerbaarheid - ken ik de bron en wie /wat is de bron - zegt dus iets over de betrouwbaarheid van de informatie/data~~. Is het bewerkte informatie zo ja in welke zin. | ii | | | | i | | |
| 29) When the information is well definied, the semantic meaning is clear to everybody and the information is in accordance with standards and definied structures. Metadata is present. There is some kind of version control in place. | | | | | | | |
| 30) ~~Being up to date and accurate~~ | i | | | | | | |
| 31) From a managerial point of view it has to be authoritative (central versions of truth need to be clear across multiple information repositories), ~~concise ('just enough', aligned with stakeholder needs) and embedded in (automated) processes in such a way (from creation to ETL operations to reporting) that the stakeholders are incentivized to keep the information accurate and up-to-date~~. From a technical point of view it has to be created transactional (in accordance to the ACID principles) and linked with other information ~~to ensure consistency. The level of information detail should also be consistent, either agreed across all stakeholders or tailored on a per stakeholder level.~~ | i | | ii | | ii | | |
| 32) ~~Is the data accurate? Is the data formatted correctly (e.g., is it consistent)? Is the data complete? In short, is the data fit for the purpose for which I need?~~ | i | | | | | | |
| 33) The information is coherent, true and can be used without explanation | ii | | | | | | |
| 34) ~~effective~~, efficient, reliable, ~~timely~~, | | | ii | ii | ii | | |
| 35) ~~Accuracy~~, ~~Completeness~~, Authenticity, Auditable, ~~Relevant~~, ~~Traceable~~ | ii | | | | | | |
| | Accuracy | Precision | Reputation | Objectivity | Traceability | Appropriate Amount | Ease of Use |

|  | Effectiveness | Completeness | Relevancy | Currency | Timeliness | Interpretability |
|---|---|---|---|---|---|---|
| 1) Dimensions to be potentially be taken into account are: ~~completeness~~, ~~accuracy, timeliness~~, validity, ~~consistency~~, conformance, ~~relevance~~, ~~interpretability~~, .. |  | i | i |  | i | i |
| 2) correct, ~~complete~~, relevant, integer, ~~timely (= data covers the period I'm interested in), available~~ |  | i | i |  | i |  |
| 3) ~~Accuracy~~, Source trustworthiness, ~~Relevancy, Completeness, Currentness~~ |  | i | i | i |  |  |
| 4) The meaning ( what the information represents) is clear, The context is clear, The Realibilty of the information is high ( know where it comes from and how it has been generated), ~~The Information is precise~~ |  |  |  |  |  |  |
| 5) ~~Accuracy~~, recency, reliability, ~~relevance~~, ~~availability/accessibility~~, ~~objectivity~~, ~~interpretability/understandability~~, coherence, ~~completeness~~ |  | i | i | ii |  | i |
| 6) ~~Complete~~, correct and minimal explaining what, why and how in a story telling manner |  | i |  |  |  |  |
| 7) ~~Relevant~~, with just the right amount of detail |  |  | i |  |  |  |
| 8) ~~Accurate~~ and time bound data, most of all fit for purpose (how good is the data for our needs). | ii |  |  |  | ii |  |
| 9) correct, ~~understandable~~, not confusing, good argumentation |  |  |  |  |  | ii |
| 10) ~~Accuracy of data~~, ~~Relevance of data/ relevance of information~~, Format/ presentation, Ease of processing, Transparency of sources used, ~~Ease of information consumption~~ |  |  | i |  |  | ii |
| 11) Was the information useful? | ii |  |  |  |  |  |
| 12) ~~Completeness (missing records/chapters)~~, |  | i |  | i |  | ii |

| | Effectiveness | Completeness | Relevancy | Currency | Timeliness | Interpretability |
|---|---|---|---|---|---|---|
| ~~timing (recent/real time vs. outdated information)~~, structure, correctness (typos), information origin/source, realistic vs. fictional information | | | | | | |
| 13) Does it make sense in the context that I want to use it? | | | | | | |
| 14) ~~Accurate, objective, relevant, complete, current, understandable~~ | | i | i | i | | |
| 15) ~~Availability~~, usability, reliability | ii | | | | | |
| 16) source is reliable, fact based analysis, verifiable | | | | | | |
| 17) ~~accurate (completeness, existence, ..), relevant, secure (incl availability, restricted access)~~ | | i | i | | | |
| 18) Reliability of the sources, ~~completeness~~, coherence. | | i | | | | |
| 19) ~~availability~~ and reliability | | | | | | |
| 20) reliable source/author, ~~recent date~~, frequency of occurrence, frequency of consultation/use, probability of correctness | | | | | i | |
| 21) Reliability & ~~accuracy~~ | | | | | | |
| 22) ~~understandable (structure, representation)~~, completeness, accuracy, | | i | | | | |
| 23) ~~Completeness~~, Correctness, ~~depending on the type: up-to-date or not? (Old information can still be relevant for evolutions etc.)~~, Trustworthiness. (Are we sure the information is untampered with? Do we know the source?) | | i | | i | | |
| 24) Information should be ~~accurate, up-to-date~~, reliable, trustworthy, protected agains undesirable modification, ~~available when needed~~, limited to what is needed, ~~easily accessible~~ and secure. | | | | | i | |
| 25) ~~The relevance of the data, Objective results~~, different sources compared | | | i | | | |
| | Effectiveness | Completeness | Relevancy | Currency | Timeliness | Interpretability |

|  | Effectiveness | Completeness | Relevancy | Currency | Timeliness | Interpretability |
|---|---|---|---|---|---|---|
| 26) Mainly the source of the information |  |  |  |  |  |  |
| 27) ~~That it's accurate across various platforms for all users (example: sales figures: same revenues, same invoiced values)~~ it is somehow cleaned - no raw data for wrong interpretation. |  |  |  |  |  |  |
| 28) ~~Traceerbaarheid - ken ik de bron en wie /wat is de bron - zegt dus iets over de betrouwbaarheid van de informatie/data~~. Is het bewerkte informatie zo ja in welke zin. |  |  |  |  |  | ii |
| 29) When the information is well definied, the semantic meaning is clear to everybody and the information is in accordance with standards and definied structures. Metadata is present. There is some kind of version control in place. |  |  |  |  |  | ii |
| 30) ~~Being up to date and accurate~~ |  |  |  | i |  |  |
| 31) From a managerial point of view it has to be authoritative (central versions of truth need to be clear across multiple information repositories), ~~concise ('just enough', aligned with stakeholder needs) and embedded in (automated) processes in such a way (from creation to ETL operations to reporting) that the stakeholders are incentivized to keep the information accurate and up to date~~. From a technical point of view it has to be created transactional (in accordance to the ACID principles) and linked with other information ~~to ensure consistency. The level of information detail should also be consistent, either agreed across all stakeholders or tailored on a per stakeholder level~~. |  |  |  | i |  |  |
| 32) ~~Is the data accurate? Is the data formatted~~ |  | i |  |  |  |  |
|  | Effectiveness | Completeness | Relevancy | Currency | Timeliness | Interpretability |

|  | Effectiveness | Completeness | Relevancy | Currency | Timeliness | Interpretability |
|---|---|---|---|---|---|---|
| ~~correctly (e.g., is it consistent)? Is the data complete? In short, is the data fit for the purpose for which I need?~~ | | | | | | |
| 33) The information is coherent, true and can be used without explanation | | | | | | |
| 34) ~~effective~~, efficient, reliable, ~~timely~~, | i | | | | i | |
| 35) ~~Accuracy, Completeness~~, Authenticity, Auditable, ~~Relevant, Traceable~~ | | | | | | |

|   | Consistency | Conciseness | Understandability | Accessibility | Availability |
|---|---|---|---|---|---|
| 1) Dimensions to be potentially be taken into account are: ~~completeness, accuracy, timeliness~~, validity, ~~consistency~~, conformance, ~~relevance, interpretability~~, .. | i |   |   |   |   |
| 2) correct, ~~complete, relevant~~, integer, ~~timely (= data covers the period I'm interested in)~~, available |   |   |   |   | i |
| 3) ~~Accuracy~~, Source trustworthiness, ~~Relevancy, Completeness, Currentness~~ |   |   |   |   |   |
| 4) The meaning (what the information represents) is clear, The context is clear, The Realibilty of the information is high (know where it comes from and how it has been generated), ~~The Information is precise~~ |   |   | ii |   |   |
| 5) ~~Accuracy~~, recency, reliability, ~~relevance~~, availability/accessibility, objectivity, interpretability/understandability, coherence, ~~completeness~~ | ii |   | i | i | i |
| 6) ~~Complete~~, correct and minimal explaining what, why and how in a story telling manner |   |   | ii |   |   |
| 7) ~~Relevant~~, with just the right amount of detail |   |   |   |   |   |
| 8) ~~Accurate~~ and time bound data, most of all fit for purpose (how good is the data for our needs). |   |   |   |   |   |
| 9) correct, ~~understandable~~, not confusing, good argumentation |   |   | i |   |   |
| 10) ~~Accuracy of data, Relevance of data/ relevance of information~~, Format/ presentation, Ease of processing, Transparency of sources used, ~~Ease of information consumption~~ | ii | ii |   |   |   |
| 11) Was the information useful? |   |   |   |   |   |
| 12) ~~Completeness (missing records/chapters), timing (recent/real-time vs. outdated information)~~, structure, correctness (typos), information origin/source, realistic vs. fictional information |   |   |   |   |   |
| 13) Does it make sense in the context that I want to use it? |   |   | ii |   |   |
| 14) ~~Accurate, objective, relevant, complete, current, understandable~~ |   |   | i |   |   |
| 15) ~~Availability~~, usability, reliability |   |   |   |   | i |
| 16) source is reliable, fact based analysis, verifiable |   |   |   |   |   |
| 17) ~~accurate (completeness, existence, ..), relevant, secure (incl availability, restricted access)~~ |   |   |   | i | i |
| 18) Reliability of the sources, ~~completeness~~, coherence. | ii |   |   |   |   |
| 19) ~~availability~~ and reliability |   |   |   |   | i |
| 20) reliable source/author, ~~recent date~~, frequency of occurrence, frequency of consultation/use, probability of correctness |   |   |   |   |   |
| 21) Reliability & ~~accuracy~~ |   |   |   |   |   |

|  | Consistency | Conciseness | Understandability | Accessibility | Availability |
|---|---|---|---|---|---|
| 22) ~~understandable (structure, representation), completeness, accuracy,~~ |  |  | i |  |  |
| 23) ~~Completeness,~~ Correctness, ~~depending on the type: up to date or not? (Old information can still be relevant for evolutions etc.),~~ Trustworthiness. (Are we sure the information is untampered with? Do we know the source?) |  |  |  |  |  |
| 24) Information should be ~~accurate, up to date~~, reliable, trustworthy, protected agains undesirable modification, ~~available when needed,~~ limited to what is needed, ~~easily accessible~~ and secure. |  |  |  | i | i |
| 25) ~~The relevance of the data, Objective results,~~ different sources compared |  |  |  |  |  |
| 26) Mainly the source of the information |  |  |  |  |  |
| 27) ~~That it's accurate across various platforms for all users (example: sales figures: same revenues, same invoiced values)~~ it is somehow cleaned - no raw data for wrong interpretation. |  |  |  |  |  |
| 28) ~~Traceerbaarheid - ken ik de bron en wie /wat is de bron - zegt dus iets over de betrouwbaarheid van de informatie/data.~~ Is het bewerkte informatie zo ja in welke zin. |  |  |  |  |  |
| 29) When the information is well defined, the semantic meaning is clear to everybody and the information is in accordance with standards and definied structures. Metadata is present. There is some kind of version control in place. |  |  |  |  |  |
| 30) ~~Being up to date and accurate~~ |  |  |  |  |  |
| 31) From a managerial point of view it has to be authoritative (central versions of truth need to be clear across multiple information repositories), ~~concise ('just enough', aligned with stakeholder needs) and embedded in (automated) processes in such a way (from creation to ETL operations to reporting) that the stakeholders are incentivized to keep the information accurate and up to date~~. From a technical point of view it has to be created transactional (in accordance to the ACID principles) and linked with other information ~~to ensure consistency. The level of information detail should also be consistent, either agreed across all stakeholders or tailored on a per stakeholder level~~. | i | i |  |  |  |
| 32) ~~Is the data accurate? Is the data formatted correctly (e.g., is it consistent)? Is the data complete? In short, is the data fit for the purpose for which I need?~~ | i |  |  |  |  |
| 33) The information is coherent, true and can be used without explanation | ii |  | ii |  |  |
| 34) ~~effective,~~ efficient, reliable, ~~timely,~~ |  |  |  |  |  |
| 35) ~~Accuracy, Completeness,~~ Authenticity, Auditable, ~~Relevant, Traceable~~ |  |  |  |  |  |

# Appendix 5 – Detailed Analysis Question 1 - Analysis Step 3

The table below contains the results of analysis step 3 as explained in the methodology section. The text in strikethrough are criteria that were mapped in analysis step 1, and the text in red are criteria that were mapped to the initial design of the IQRM artefact in analysis step 2. The remaining parts of the respondent's answers are discussed, and disposition is decided, with proper justification.

| Respondent's answers | Disposition | Justification |
|---|---|---|
| 1) *Fitness for use, so depending on the context and the intended usage, the criteria can be different and threshold for being considered as qualitative might vary. Dimensions to be potentially be taken into account are: ~~completeness, accuracy, timeliness,~~ validity, ~~consistency,~~ conformance, ~~relevance, interpretability~~, ..* | Not added to model | 'Fitness for use' is a combination of effectiveness and relevancy, so it can't be a separate criterion. One could even argue that it also is an overall description of quality, so again it does not qualify as a separate criterion. 'Conformance' is linked to compliance, which is not a quality issue but rather a property on how information is used according to stated rules. |
| 16) *source is reliable, fact based analysis, verifiable* | Not added to model | 'fact based analysis' is on our view not a criterion per se but rather a means to achieve some of the criteria, e.g. accuracy. It can also be argued that it is equivalent to objectivity. |
| 20) *reliable source/author, ~~recent date~~, frequency of occurrence, frequency of consultation/use, probability of correctness* | Not added to model | 'frequency of consultation' is a measure on how much the information is used, not an information quality criterion |
| 24) *Information should be ~~accurate, up-to-date~~, reliable, trustworthy, protected against undesirable modification, ~~available when needed~~, limited to what is needed, ~~easily accessible~~ and secure.* | Not added to model | 'secure' is covered by the criteria 'accessibility' and 'availability', as well as 'completeness' and 'accuracy'. |
| 29) *When the information is well defined, the semantic meaning is clear to everybody and the information is in accordance with standards and definied structures. Metadata is present. There is some kind of version control in place.* | Not added to model | 'Version control' is a means to protect some of the quality criteria, but not a criterion in itself. |
| 31) *From a managerial point of view it has to be authoritative (central versions of truth need to be clear across multiple information repositories), ~~concise ('just enough', aligned with stakeholder needs) and embedded in (automated) processes in such a way (from creation to ETL operations to reporting) that the stakeholders are incentivized to keep the information accurate and up-to-date~~. From* | Not added to model | 'created transactional' is seen to be more a means to ensure some of the other criteria in the IQRM. |

| Respondent's answers | Disposition | Justification |
|---|---|---|
| *a technical point of view it has to be created transactional (in accordance to the ACID principles) and linked with other information* ~~to ensure consistency. The level of information detail should also be consistent, either agreed across all stakeholders or tailored on a per stakeholder level~~. | | |
| 34) ~~effective~~, *efficient*, reliable, ~~timely,~~ | Not added to model | Efficiency can relate to how effective information is, hence we consider it as equivalent to that criterion. |
| 35) ~~Accuracy, Completeness,~~ Authenticity, *Auditable,* ~~Relevant, Traceable~~ | Not added to model | Auditable is about the possibility to independently assess not only the accuracy but in essence any of the quality criteria included in the draft IQRM. Hence, we believe this is not a quality criterion but rather a sort of meta-property of the information. When interpreting auditability somewhat narrower, it could also be argued that traceability and auditable are more or less equivalent. |

**Appendix 6 – Survey on Information Quality Criteria validation – Detailed remarks Question 2**

The following comments were given to this question:

- All 18 criteria are relevant and depending on the project at hand some are more important than others.

- The relevance of the criteria depends on the context and the use-case. E.g. Time-Series information for real-time decision making in e.g. Industrial Asset Mgmt. context has other key criteria than e.g. information about customer churn or information that represents an ontology.

- The definitions for 'Currency' and 'Timeliness' seem duplicates in my opinion, but they are both different criteria. So I would change the definition for Timeliness: "Timeliness refers to the time expectation for accessibility and availability of information. Timeliness can be measured as the time between when information is expected and when it is readily available for use."

- I think in the end they are all relevant, depending on the intended usage. Accessibility works in 2 directions: those who should have access and those who should not, and there is a difference to be made (also applicable to other criteria) between the raw data and derivates

- The 'quickly' in the definition of effectiveness is confusing, as it also hints at efficiency. Information that requires some thought, processing, effort, time, etc. to meet the task can still be effective in my opinion (but maybe its efficiency can be improved).

- Currency vs. timeliness --> difference(s)? Accessibility & availability do not impact the quality IMO

- All of above could be relevant, but I tried to tag those that I believe are the most relevant.
- as defined, appropriateness and completeness aims at the same depending on the definition interpretation precision can also cover accuracy... where in the given definition precision is not looking at the content but the relationship between the different parts of the data creating the information.. there is a distinction between quality of the information (and its components) and the dimensions that generate access to information... when not having the information access and availability are relevant, however this starts from the intake that you know that there is information which is not available or accessible... when in the situation of having access to that information it is not a relevant parameter since you have the information.... as such the perspective will impact the relevance of different parameters. In general it is impossible to judge or assess on the quality of information if it is not available.. since you do not have the "information" to make that assessment.
- Some of them refer to the same aspect: (i) appropriate amount (ii) completeness. (i) Accessibility and (ii) availability. Availability and accessibility lead to understandability. There is a means and a target. Both are in a different level
- Timeliness and currency might be somewhat redundant.
- I don't agree only with the "objectivity" if the information is accurate, precise and traceable - it should up to the user to interpret it objectively. All the others look fair - I would still have a question mark about reputation
- Some criteria will become more and less relevant in relation to the task at hand.
- I associated Accessibility with Accessibility legislation (e.g. Web Accessibility Directive), I associate usability of information with both understandability and interpretability. Ease of use + appropriate amount also seems to relate to it. For interpretability I associate it with a common set of abstractions or codification. Effectiveness, ease of use and appropriate amount also seem to relate to each other for me on the topic of just enough information towards target audiences. Towards technical needs I seem to miss atomicity ("undividedness") and durability (which is a huge buzzword in distributed ledgers nowadays such as blockchain, and not sufficiently covered by traceability i.m.h.o.)

# Appendix 7 – Survey on Information Quality Criteria validation – Detailed answers Question 3

The following was mentioned by the respondents that completed the question on potential contingencies (N=29)

| Resp | Answers to question #3 |
|---|---|
| 1 | Traceability in an audit environment<br>Objectivity in a decision-making context<br>Precision in a space-rocket-building context |
| 2 | Volume can be really important when training a machine learning model<br>Interpretability is in most cases important across the process of correctly collecting data, processing the data and interpreting the data and products derived from the data<br>Traceability is often essential in the context of compliance and regulatory reporting, but also to build transparency and trustworthiness in the products created from the data |
| 3 | Objectivity depends on the task. If you want opinions, then subjective information can be qualitative. |
| 4 | Conciseness is important if the processing (human or non-human) depends strongly on the conciseness of the information<br>Timeliness is more important for information where decisions might be strongly influenced by it |
| 5 | fraud detection --> recency, accuracy, reliability, availability, ...<br>Web Application Vulnerability Assessment --> timeliness, accessibility, interpretability, currency, completeness, traceability, reputation, ... |
| 6 | Accuracy is important when the butterfly effect is at play.<br>Reputation is more important when marketing is at play.<br>Objectivity is important when politics are an issue for credibility.<br>Currency is essential in a fast-paced world |
| 7 | Healthcare<br>Security |
| 8 | Understandability is strongly dependent on target audience for which information is intended.<br>Conciseness is relevant for summaries but not for long reads/ whitepapers.<br>Objectivity is relevant for scientific information but not for opinions |
| 9 | Required precision of values might depend on the type of calculation or use. When I buy bread and the sack mentions 600 grams, well that is sufficient for me, even when the actual weight is 606,66 gram.<br>News items are rarely objective, still we all watch or listen to news broadcasts. Luckily, we can make up our own mind in the end.<br>The weather forecast for today is sunny and cloudless. I am not that much interested in who actually provided the forecast or which data was actually used (reputation, traceability)<br>Mostly I don't care who else uses the information that I use (accessibility) |
| 10 | Consistency is more important when comparing different sets of the same information type (e.g. comparing sensor observations from multiple sensors of the same type for a specific time interval)<br>Understandability is not relevant for machine-readable information.<br>Accuracy of information is more important when monitoring a nuclear reactor's statistics, compared to measuring the amount of people in a crowd/shopping street |
| 11 | Legitimate business interest, related to privacy.<br>Criticality of the information. Is it a lifesaving need or routine need? This could be linked to timeliness |
| 12 | reputation may be more important in important decision-making situations.<br>conciseness may be more important if analysis is not or less automated.<br>understandability may be less important if information is presented to experts |
| 13 | as stated, the CSF or KPI that determine the quality of data all depend on the purpose and the conditions in which information is (to be) generated<br>warning and alarm signals on fixed and movable assets (mechanic / IoT / phones etc..) are to be timely / accessible / easily interpretable etc. if not you run into the "risk" of living the externality that might occur as a consequence of the unwanted event (or alike)<br>the publication of legislation and depository of authentic acts etc. generate certainty / a fixed date and create the legal substance which is stipulated in the text. interpretability / completeness / accessibility / etc. |
| 14 | Utilization of products. Reading the manual always help.<br>Analysis of risk: dashboard. |
| 15 | accessibility is particularly important for personal data |

| Resp | Answers to question #3 |
|---|---|
| 16 | objectivity may be less important when reading/writing personal letters, when conducting personal (sometimes emotional) conversations<br>completeness cay be unnecessary when the purpose is to get acquainted with a certain subject<br>conciseness should not be the aim of a novelist (fiction) |
| 17 | Results of a blood sample need to be accurate to decide on the appropriate medication |
| 18 | Completeness is more important for a purchase order than an application manual<br>Accuracy is more important for a tender compared to a company newsletter<br>Conciseness is more important for a project dashboard compared to a account transaction history |
| 19 | Privacy<br>Cyber Security |
| 20 | Real time transactions<br>For students<br>For a business case<br>For the law |
| 21 | Understandability depends on the goal of the information (e.g. engineering data for engineers or for everybody) |
| 22 | for data science: raw data is key - have it accessible, traceable, correct - it's fine.<br>marketing & sales: understandability, relevant, ease of use. |
| 23 | Scientific report on the quality of water (accuracy > timeless)<br>Accounting report (completeness > ease of use) |
| 24 | Scientific measurements : precision, traceability and accuracy are very important<br>websites: completeness is less important, conciseness and understandability are very relevant<br>buying goods: reputation, accuracy are very important |
| 25 | Accuracy definitely more important in transactional data |
| 26 | Accessibility of draft information versus final information (policies, reporting, decisions)<br>Accuracy, Objectivity, Reputation, Relevancy, Timeliness, Consistency, Conciseness, Understandability and Accessibility of an assessment report<br>Atomicity, understandability, conciseness, interpretability, timeliness, currency, relevance, appropriate amount and accuracy of (EA) IT architectural information<br>The availability of software (algorithms and code) used to recover backup formats alongside the backups themselves<br>Interpretability of high assurance software code vs. high performance code |
| 27 | For some things, data quality must be part of a large measurable data set (e.g., having enough data to accurately program machine learning with no bias (robust data set needed))<br>Accuracy is an important quality for financial reporting (e.g., filing a tax return)<br>The format of data can be extremely important when coding (e.g., creating unit tests for software) |
| 28 | Availability of departure data for trains and flights are more important<br>Correctness of price data<br>Precision for the time data about delivery of goods |
| 29 | Objectivity: when analyzing customer preferences, objectivity is not an indicator.<br>Consistency: when comparing cases, the format should be identical. But to me, the consistency does not imply a notion of information quality, but rather a notion of ease for analysis.<br>Timeliness: very important in real time monitoring (e.g. Healthcare or stock exchange)<br>Precision: again, in healthcare, precision can be very crucial in some cases, but less crucial in other cases. Same goes for production of equipment etc.<br>Reputation: for navigation purposes (e.g. aircrafts), the source of information is quite crucial.<br>Completeness: when drawing conclusions that have a severe impact (e.g. policy making), the information set used should be broad enough to draw conclusions. |

# Appendix 8 – Survey on Information Quality Criteria validation – Detailed answers Question 5

This appendix contains the detailed answers received on question 5 on potential additional information quality criteria.

- Respondent 1
    - Credibility: is the information coming from a trustworthy source (partially covered), Processability: is the data machine-readable
- Respondent 2
    - Noise Factor, Bias Factor
- Respondent 2:
    - Objectivity, Comparability is the extent to which data from one source can be compared directly to another (historic) source
- Respondent 4
    - Multiple sources confirming correctness, Source of funding, Grading of surveyors
- Respondent 5
    - Not really, but you might want to hold it against the SABSA business attributes (https://onlinelibrary.wiley.com/doi/pdf/10.1002/9780470476017.app1). They might offers some inspiration, e.g. a metric type (soft vs hard) or suggested measurement approach
- Respondent 6
    - Usefulness, which is like relevancy, but more like 'after the fact'. When I have to perform a task that requires some information, and I get information, I can evaluate its relevancy. After the task is performed, I can also evaluate

whether the information was useful. Evaluating usefulness before performing the task is difficult; would be more like guessing or assessing.

- Respondent 7
    - clarity on the purpose of the information
    - clarity on the context in which the information is gathered and communicated
- Respondent 8
    - Confidentiality
- Respondent 9
    - Durability
    - Link ability (privacy, ability to link information of interest)
    - Detectability (privacy, ability to distinguish whether information exists or not)

| Suggested Criterion | Discussion | Disposition |
| --- | --- | --- |
| Credibility | Equivalent to the combined criteria 'Reputation' and 'Objectivity' | Not Added |
| Processability | Equivalent to the combined criteria 'Ease of Use' and 'Interpretability' | Not Added |
| Noise Factor | Equivalent to the combined criteria 'Accuracy' and 'Precision' | Not Added |
| Bias Factor | Equivalent to the criterion 'Objectivity' | Not Added |
| Objectivity | Exists already as criterion in the model | Not Added |
| Comparability | Equivalent to the criterion 'Traceability' | Not Added |
| Multiple sources confirming correctness | Equivalent to the combined criteria 'Traceability', 'Reputation' and 'Objectivity' | Not Added |
| Source of funding | Equivalent to the criterion 'traceability' | Not Added |
| Grading of surveyors | Not sure about what the respondent meant | Not Added |
| Usefulness | The suggestion describes a true situation, but we believe it is not relevant in our context; this model is about criteria for quality and is agnostic to when measurement or assessment takes place, and we consider useful as an aggregation of several of the stated criteria, i.e. | Not Added |
| Clarity on the purpose of the information | This is a valid concern about the use of information, but it is not a quality criterion, rather a property of the information user | Not Added |
| Clarity on the context in which the information is | Equivalent to the combined criteria 'Traceability' and 'Reputation' | Not Added |

| Suggested Criterion | Discussion | Disposition |
|---|---|---|
| gathered and communicated | | |
| Confidentiality | Confidentiality is related to the way the information is used but is not a quality criterion in itself. Other criteria like 'Accessibility' will help to achieve confidentiality should that be required. | Not Added |
| Durability | We understand this criterion might be relevant for distributed ledgers, but we think this requirement can be covered by the combined criteria of 'Availability' and 'Traceability' | Not Added |
| Link ability (privacy, ability to link information of interest) | Linking or associating information items with other information items is indeed a thing but is more an issue on how information is used than it is a quality criterion. | Not Added |
| Detectability (privacy, ability to distinguish whether information exists or not) | We consider the detectability criterion equivalent to 'Accessibility'. Furthermore, we assume that considering information quality criteria is only meaningful when the information exists. | Not Added |

## Appendix 9 – Comparison with other industry Frameworks / standards

This appendix contains a mapping between the attributes of our proposed IQRM and the quality dimensions defined in the Data Management Wiki (Source: https://www.dama-nl.org/wp-content/uploads/2020/09/DDQ-Dimensions-of-Data-Quality-Research-Paper-version-1.2-d.d.-3-Sept-2020.pdf)

| IQRM Attribute | Data Management dimension |
|---|---|
| ACCURACY | Accuracy |
| PRECISION | Precision |
| SOURCE RELIABILITY | Reputation<br>Reliability<br>Credibility<br>Plausibility |
| OBJECTIVITY | Objectivity |
| TRACEABILITY | Traceability |
| APPROPRIATE AMOUNT | Appropriateness |
| USABILITY & EFFECTIVENESS | Clarity |
| COMPLETENESS | Completeness |
| RELEVANCY | Relevance |
| CURRENCY | Currency |
| TIMELINESS | Timeliness |
| INTERPRETABILITY | |
| CONSISTENCY | Coherence<br>Consistency |
| UNDERSTANDABILITY | |
| ACCESSIBILITY | Accessibility<br>Obtainability |
| AVAILABILITY | Recoverability<br>Findability<br>Availability |
| | Ability to represent null values |
| | Metadata compliance |
| | Comparability |
| | Compliance |
| | Confidentiality |
| | Equivalence |
| | Granularity |
| | Integrity |
| | Latency |
| | Link ability |
| | Naturalness |
| | Portability |
| | Punctuality |
| | Reasonability |
| | Redundancy |
| | Referential integrity |
| | Reproducibility |
| | Retention period |
| | Uniqueness |
| | Validity |
| | Valuability |
| | Variety |
| | Volatility |